\documentclass[a4paper,11pt]{article}
\pdfoutput=1 

\usepackage{jcappub} 

\usepackage[T1]{fontenc} 

\usepackage{journalsbib}
\usepackage{comment}
\usepackage{soul}
\usepackage{xcolor}
\usepackage{multirow}
\usepackage{verbatim}
\usepackage{enumitem}
\subheader{ULB-TH/25-09}

\title{\boldmath 21 cm Cosmology Sensitivity to Small-Scale Structure: Warm vs Neutrino-Interacting Dark Matter}

\author[a]{V. Dandoy,}
\author[a]{C. D\"oring,}
\author[a]{G. Facchinetti,}
\author[a, b]{L. Lopez-Honorez,}
\author[a]{J. R. Schwagereit,}

\affiliation[a]{Service de Physique Th\'{e}orique \& Brussels Laboratory of the Universe BLU-ULB, CP 225, Universit\'{e} Libre de Bruxelles,  Boulevard du Triomphe, B-1050 Brussels, Belgium}

\affiliation[b]{Theoretische Natuurkunde \& The International Solvay Institutes, \\
Vrije Universiteit Brussel, Pleinlaan 2, B-1050 Brussels, Belgium}

\emailAdd{justus.roman.schwagereit@ulb.be}
\emailAdd{gaetan.facchinetti@ulb.be}
\emailAdd{llopezho@ulb.be}
\emailAdd{virgile1997@hotmail.com}
\emailAdd{christian.doering1990@gmx.de}

\abstract{The $21\,$cm signal originating from Cosmic Dawn to the Epoch of Reionisation is highly sensitive to the processes governing star formation in the early universe as well as new physics. In this work, we focus on the imprint of non-cold dark matter (DM), which impacts the formation of the smallest halos. Our goal in particular is to clarify whether near-future radio telescopes such as the Hydrogen Epoch of Reionisation Array (HERA), will be able to distinguish between free-streaming dark matter, specifically in the form of thermal warm DM (WDM), and collisional damping due to neutrino-DM ($\nu$DM) interactions giving rise to larger overdensities on small scales. 
For that purpose we first implement a mapping between the two models in terms of a cutoff scale and determine detection thresholds for the two DM models. Using Fisher matrix forecasts,
we show that $\nudm$ interaction strengths down to $\sigma_\nudm\sim 3\times10^{-35}$ cm$^2$ could be probed by $21\,$cm cosmology  when considering two populations of galaxies for a GeV mass DM. This would allow to either confirm or rule out a recent claimed preference for a non-zero $\nudm$ interaction in Lyman-$\alpha$ data. Furthermore, we find that HERA will not be able to distinguish between $\nudm$ and WDM. In the latter context, the threshold for detection of $\nudm$ interactions translates into WDM with mass up to $m_{\rm WDM}\sim 9$ keV that could be detected by HERA.
}

\newcommand{\lcut}{\lambda_{\rm cut}}
\newcommand{\lcutfid}{\lambda_{\rm cut}^{\rm fid}}

\newcommand{\nudm}{{\nu{\rm DM}}}

\newcommand{\btheta}{\ensuremath{\boldsymbol \theta}}
\newcommand{\brho}{\ensuremath{\boldsymbol \rho}}


\begin{document}
\maketitle
\flushbottom

\section{Introduction }
\label{sec:intro}

\noindent
With the unprecedented volume and redshift range that will soon be accessible, the $21\,$cm signal -- arising from redshifted $21\,$cm line emissions in atomic hydrogen and detectable by radio telescopes on Earth -- is expected to provide a new and unparalleled probe of our universe at the Epoch of Reionisation (EoR, $z \sim 5$) and Cosmic Dawn (CD, $z \sim 10-20$)~\cite{Pritchard:2011xb}. The current generation of interferometers, such as the Hydrogen Epoch of Reionisation Array (HERA), now fully deployed \cite{HERA:2021bsv}, is expected in the near future to detect fluctuations in the $21\,$cm signal through its power spectrum, across the redshift range $z \sim 5$–20. Preliminary results from HERA, based on data from 71 of its 331 antennas collected over 94 nights, have already provided the most stringent upper limits to date on the $21\,$cm power spectrum at $z = 8$ and $z = 10$ \cite{HERA:2022wmy}. These observations will open a new window on a largely unexplored era of cosmic history, when the first dark matter (DM) halos were forming.

DM halos are a consequence of the growth and collapse of overdensities in the matter distribution. Acting as gravitational potential wells, they drive the formation of galaxies and their subsequent effect onto the late-time universe. Therefore, as $21\,$cm cosmology experiments will probe the intergalactic medium (IGM), they will additionally indirectly probe the fundamental properties of DM that can affect the halo distribution. Here, we focus on DM properties that suppress structure formation on small scales. In particular, thermal warm dark matter (WDM), produced through relativistic freeze-out in the early universe, can free-stream out of the shallow potential wells of the tiniest overdensities due to its non-negligible velocity dispersion at the time of structure formation. As such, WDM represents the standard benchmark for non-cold dark matter (NCDM), and in particular of free-streaming NCDM.\footnote{WDM has been long proposed as a solution to some of the small-scale discrepancies between cold-DM-only simulations and observations of dwarf galaxies \cite{Bullock:2017xww, Dubinski:1991bm, Flores:1994gz, Bode:2000gq}. Although today the addition of baryonic feedback effects to the simulations alleviates the tensions \cite{Pontzen:2011ty, Madau:2014ija, Chan:2015tna, Onorbe:2015ija, Read:2015sta, Navarro:2016bfs, Paranjape:2021uyu}, the WDM model remains a plausible and interesting scenario due to its simplicity and minimality. Yet, if WDM with a -- still allowed -- mass of a few keV is to account for the entirety of the DM in the universe, the introduction of $\mathcal{O}(1000)$ extra relativistic degrees of freedom is required at the time of production.} DM could also suppress structures at small scales due to interactions with light degrees of freedom. Pressure forces can indeed further dampen the small-scale fluctuations of the matter density field \cite{Weinberg:1971mx, Boehm:2000gq, Boehm:2001hm, Boehm:2004th, Loeb:2005pm} (similarly to the Silk damping on baryons  \cite{Silk:1967kq}). This effect is referred to as collisional damping. A variety of such scenarios has been investigated in the literature. These include interactions
of dark matter with  photons~\cite{Escudero:2018thh, Boehm:2000gq, Boehm:2001hm, Boehm:2004th, Dolgov:2013una, Cyr-Racine:2012tfp, Sigurdson:2004zp, Wilkinson:2013kia, Boehm:2014vja, Schewtschenko:2014fca, Schewtschenko:2015rno, Escudero:2015yka, McDermott:2010pa, Diacoumis:2017hff, Ali-Haimoud:2015pwa, Stadler:2018jin, Lopez-Honorez:2018ipk} as well as with neutrinos~\cite{Boehm:2000gq, Boehm:2001hm, Boehm:2004th, Ali-Haimoud:2015pwa, Mangano:2006mp, Serra:2009uu, Wilkinson:2014ksa, DiValentino:2017oaw, Stadler:2019dii, Mosbech:2020ahp, Hooper:2021rjc, Akita:2023yga}, and more recently with a dark sector radiation
species~\cite{Kaplan:2009de, Das:2010ts, Diamanti:2012tg, Buen-Abad:2015ova, Lesgourgues:2015wza, Das:2017nub, Ko:2017uyb, Escudero:2018thh, Archidiacono:2019wdp, Bringmann:2006mu, Plombat:2024kla, Munoz:2019hjh, Verwohlt:2024efh}.  Probing the abundance and distribution of small-scale structures provides a powerful means to constrain the fundamental nature of dark matter. In this paper, we investigate whether a near-future radio telescope array that probes the fluctuations in the $21\,$cm signal could help to discriminate between collisionless and collisional damping, focusing on two concrete scenarios: thermal WDM and neutrinos-dark matter interactions (referred to as $\nudm$ in the following). Here we consider the configuration of the HERA telescope for our forecasts.

On the astrophysics side, efficient galaxy formation scenarios involving molecular cooling give rise to the earliest luminous objects -- responsible for initiating CD at redshifts $z > 10$ -- which could form in halos with masses as low as $10^{5-6}~{\rm M_\odot}$  at $z\sim 15$ \cite{Tegmark:1996yt, Munoz:2021psm}. These are referred to as molecular cooling galaxies (MCGs), while more massive halos which allow atomic cooling, host atomic cooling galaxies (ACGs)  with mass greater than $10^8 {\rm M_{\odot}}$. The abundance of MCGs, as well as the smallest ACGs, depends on the small-scale matter power spectrum and thus on the properties of DM. Conversely, the sensitivity of $21\,$cm cosmology towards the properties of DM depends on the properties of the first CD galaxies. In this paper, we will consider two populations of galaxies (ACGs and MCGs) and highlight how their specific properties can affect the reconstruction of the DM properties and vice-versa. 

Lyman-$\alpha$ forest observations set a stringent lower bound on the WDM mass \cite{Viel:2005qj, Viel:2013fqw} of
\begin{equation}
    m_{\rm WDM}^{{\rm Ly}\alpha}> 5.7\, {\rm keV}
    \label{eq:mWDMly}
\end{equation}
at 95\% confidence level, see Ref.~\cite{Irsic:2023equ}. Other recent constraints on WDM from other probes can also be found in e.g.~\cite{Keeley:2024brx,Tan:2024cek,Gilman:2025fhy} and references therein. Here we use the exclusion of WDM with masses $m_{\rm WDM}\lesssim5.7\,\rm keV$ at 95\% CL from~\cite{Irsic:2023equ} as a reference. Simultaneously, for the case of neutrino-DM interactions that are governed by a temperature-independent scattering cross-section $\sigma_\nudm$, an upper limit on the interaction strength of $\sigma_\nudm \lesssim 10^{-33}$ cm$^2$ for GeV DM was obtained in~\cite{Wilkinson:2014ksa} using Lyman-$\alpha$ data and assuming massless neutrinos. This bound was updated in~\cite{Mosbech:2020ahp} using CMB and SDSS BAO data and considering massive neutrinos. The presence of massive neutrinos led to a looser bound of $\sigma_\nudm \lesssim 2.2 \times 10^{-30}$ cm$^2$ for GeV-mass DM. Furthermore, using Lyman-$\alpha$ flux power spectrum data, Ref.~\cite{Hooper:2021rjc} reported a preference for a non-zero interaction strength of $\sigma_\nudm \sim 3.7 \times 10^{-32}$ cm$^2$ for GeV mass  DM. Some previous analyses have already investigated the ability of $21\,$cm data to help improve constraints on the NCDM models we consider here, namely~\cite{Giri:2022nxq, Decant:2024bpg,Verwohlt:2024efh,Schosser:2024aic,Hibbard:2022sng,Schneider:2018xba,Park:2025phj} for WDM and~\cite{Mosbech:2022uud,Dey:2022ini} for $\nudm$. 
Our analysis differs from those either in that we focus on the $21\,$cm power spectrum and not the global signal, in the implementation of the NCDM imprint onto the $21\,$cm signal, or in the modelling of the astrophysics in that we consider two galaxy populations. Our approach to the NCDM implementation is most similar to that of e.g.~\cite{Decant:2024bpg,Verwohlt:2024efh} in that we use a sharp-k cutoff in the calculation of the halo mass function. In this work, we perform a statistical analysis by means of Fisher matrix forecasts to assess the sensitivity of HERA to both scenarios and determine whether it would be possible to distinguish between them if a signal was detected. To that end, we give thorough attention to the description of the halo population and the computation of the halo mass function in scenarios with a damped matter power spectrum. Note, however, that~\cite{Mosbech:2022uud}, who focused on the $21\,$cm signal at lower redshift (with the SKA experiment), already suggested  that WDM might closely mimic $\nudm$ for $z\lesssim 10$, which would make it impossible to distinguish between those scenarios observationally.

The paper is organised as follows. In section~\ref{sec:ncdm}, we introduce the WDM and $\nudm$ models and examine and compare the linear matter power spectra obtained in both cases. In section~\ref{sec:HMF}, we relate the power spectrum to the halo distribution and, subsequently, to the astrophysical sources imprinted on the $21\,$cm signal. The astrophysical modelling is more specifically addressed in section~\ref{sec:21cm}. We discuss our methodology and results in sections~\ref{sec:analysis} and~\ref{sec:results} respectively. Finally, we conclude in section~\ref{sec:conclusion}. Further technical details are given in the appendices~\ref{app:cosmo} to~\ref{app:fisher}. 

\section{NCDM scenarios and linear perturbations}
\label{sec:ncdm}

In this work, we focus on the WDM and $\nudm$ scenarios which suppress the linear matter power spectrum on small scales and hence impact the $21\,$cm power spectrum by reducing the population of star-forming halos during cosmic dawn. This section is devoted to the different damping mechanisms and the introduction of a conveniently defined transfer function. The impact on the halo distribution and on the evolution of the IGM will be addressed in the next two sections.

\subsection{Thermal warm dark matter}

\begin{figure}[t]
    \centering
    \includegraphics[width=0.99\linewidth]{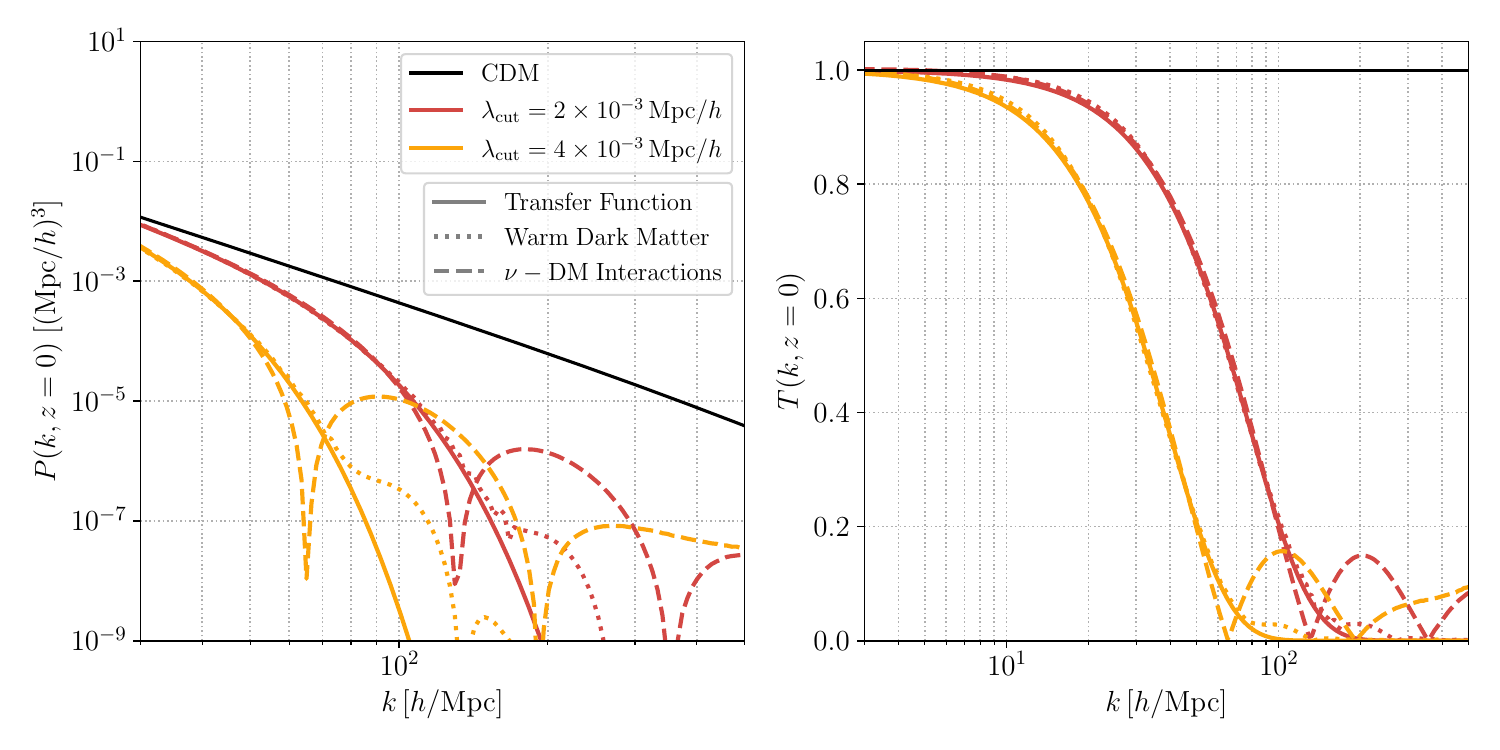}
    \caption{\textbf{Left:} The matter power spectrum at $z=0$ as a function of the comoving wavenumber. The $\Lambda{\rm CDM}+3 m_\nu$ scenario is shown as the black solid line for reference. The matter power spectra of each of the three scenarios that we consider, WDM, $\nudm$ and the fitted transfer function, are shown as dotted, dashed and solid lines respectively for two values of the cutoff scale $\lcut=2\times10^{-3}\,{\rm Mpc}/h$ (red) and $\lcut=4\times10^{-3}\,{\rm Mpc}/h$ (yellow). The corresponding $m_{\rm WDM}$ and $u_\nudm$ are shown in Table~\ref{tab:lcut_conversion}. \textbf{Right:} The square root of the ratios between the matter power spectra of the three scenarios shown in the left plot and the $\Lambda{\rm CDM}+3m_\nu$ matter power spectrum. This corresponds to the transfer functions as defined in Eq.~(\ref{eq:transfer_funtion_definition}) for $i=$ WDM (dotted) and $\nudm$ (dashed)  as well as the analytical expression for the transfer function given in Eq.~(\ref{eq:transf_function_fit_form}) (solid). }
    \label{fig:MPS&TF-LCDM+3m_nu}
\end{figure}

WDM suppresses structure formation on small scales through \textit{collisionless damping} of the perturbations. Particles that kinetically decouple from the thermal bath homogenise the cosmic fluid by streaming from overdense to underdense regions. The thermal velocity of the particles determines the comoving free streaming length $\lambda_{\textrm{fs}}$, below which perturbations are damped, defined as:
\begin{equation}
    \lambda_{\rm fs} \equiv  \int_{a_{\rm kd}}^a \frac{{\rm d}a'}{a'^2H(a')}\frac{p(a')}{\sqrt{m^2 + p^2(a')}}\,,
    \label{eq:lamfs}
\end{equation}
i.e. the comoving distance travelled by the DM since kinetic decoupling (kd). This corresponds to the time at which the elastic scattering rate between the DM particles and the species populating the thermal plasma becomes smaller than the expansion rate.
 The corresponding comoving wave number $k_{\textrm{fs}}=2\pi/\lambda_{\textrm{fs}}$ depends mainly on the mass of the particle, $m_{\textrm{WDM}}$. As shown in appendix~\ref{app:free_stream},
(see also e.g. Ref.~\cite{Viel:2013fqw} for a fit on numerical simulations), it is approximately given by
\begin{equation}
    k_{\rm fs} \simeq 4.1 \frac{r}{1+0.15\ln r } ~ h / {\rm Mpc} \quad {\rm with} \quad r \equiv \left(\frac{m_{\rm WDM}}{{\rm keV}}\right)\left(\frac{T_{ \gamma, 0}}{T_{\rm d, 0}}\right) \, ,
    \label{eq:k_free_streaming}
\end{equation}
where $T_{\rm \gamma, 0}$ is the photon temperature today and $T_{\textrm{d}, 0} \equiv T_{\rm kd}a_{\rm kd}$ is the decoupling temperature today (product of the temperature and scale factor at decoupling). In the vanilla fermionic WDM model, the latter can be written \cite{Viel:2005qj},
\begin{equation}
\begin{split}
    T_{\rm d, 0} & = \left( \frac{4\pi^2}{3\zeta(3)} \frac{0.12}{g_{\rm DM}} \frac{\rho_{\rm c, 0}}{T_{\gamma, 0}}\frac{1}{\rm keV} \right)^{1/3} \left( \frac{\Omega_{\rm DM} h^2}{0.12} \frac{\rm keV}{m_{\rm WDM}}\right)^{1/3} T_{\gamma, 0}\\
    & \simeq 0.160 \left( \frac{\Omega_{\rm DM} h^2}{0.12} \frac{\rm keV}{m_{\rm WDM}}\right)^{1/3} T_{\gamma, 0}\,,
    \end{split}
    \label{eq:Td0}
\end{equation}
where $\rho_{\rm c, 0}$ is the critical density of the Universe today and the number of DM degrees of freedom is taken to be $g_{\rm DM} = 2$. The free-streaming wavenumber of a 1 keV DM particle is thus $k_{\rm fs} \simeq 20 ~h/$Mpc. 
For comparison, the free-streaming scale of  a cold dark matter (CDM) candidate with mass $m_{\textrm{WDM}}=10\,\textrm{GeV}$ that decoupled thermally at a temperature $T_{\rm kd}=100\,\textrm{MeV}$ is $k_{\textrm{fs}} \sim 10^6\,h/\textrm{Mpc}$ \cite{Hofmann:2001bi, Green:2005fa, Bertschinger:2006nq,  Bringmann:2006mu} -- five orders of magnitude larger. The effect of WDM free-streaming on the matter power spectrum, as computed by the public Boltzmann solver code {\tt CLASS}~\cite{Blas:2011rf, Lesgourgues:2011rh}, is shown in the dotted lines in the left panel of Fig.~\ref{fig:MPS&TF-LCDM+3m_nu} for $m_{\rm WDM}\sim 5.1$ (red) and $2.8$ keV (yellow) in red. It can be seen that the WDM additionally generates acoustic oscillations at scales $k>2k_{\textrm{fs}}$ and a dark integrated Sachs-Wolfe effect \cite{Boyanovsky:2010pw}. Nonetheless, as shown in the figure, these effects are minor in comparison to the sharp free-streaming cutoff. In Fig.~\ref{fig:MPS&TF-LCDM+3m_nu}, the case of the CDM and $\nudm$ are  shown for comparison with continuous black line and coloured dashed lines respectively, see discussion below.

\subsection{Neutrino - dark matter interactions}

The pressure induced by neutrino-dark matter interactions balances gravitational forces in the primordial plasma and prevents the growth of overdensities. This process is similar to the photon-baryon scattering which kept the baryon density fluctuations oscillating before decoupling and prevented them from collapsing gravitationally. This results in a suppression of the matter power spectrum that is referred to as \textit{collisional damping}. More precisely, all modes inside the horizon until dark matter-neutrino decoupling or, said differently, below the diffusion scale set by the interaction efficiency, are damped. For a temperature-independent scattering cross-section, the interaction strength can be  characterised by \cite{Boehm:2001hm} 
\begin{equation}
    u_{\nu \rm DM} = \frac{\sigma_{\nu \rm DM}}{\sigma_{\rm Th.}}\left(\frac{m_{\rm DM}}{100 \, \rm GeV}\right)^{-1},
\end{equation}
where $\sigma_{\nu \rm DM}$ and $\sigma_{\rm Th.}$ are the DM-neutrino scattering and Thomson cross-sections, respectively while $m_{\rm DM}$ refers to the mass of the interacting DM (which is {\it not} assumed to be WDM). A GeV mass candidate with a $\nudm$  interaction strength of $u_{\rm \nu DM}=3.3 \times 10^{-4}$, excluded by the most recent CMB+BAO bound~\cite{Mosbech:2020ahp} when considering massive neutrinos, corresponds thus to a scattering cross-section of $ \sigma_\nudm= 2.2 \times 10^{-30}$ cm$^2$ for a GeV DM candidate, as reported in Sec.~\ref{sec:intro}.
Notice that throughout this paper,  we assume that all three Standard Model (SM) neutrinos are massive with the same mass $m_\nu = 0.02~{\rm eV}$ when considering $\Lambda{\rm CDM}+3 m_\nu$ (CDM for short), WDM and $\nudm$ cosmologies.

Interestingly, beyond a mere damping of the matter power spectrum on small scales, the competition between pressure and gravity further leads to the propagation of  dark acoustic oscillations, similar to the baryon acoustic oscillations. These leave a sizeable imprint at a scale around the sound horizon of neutrinos at decoupling that is clearly visible in the dashed lines in both panels of Fig.~\ref{fig:MPS&TF-LCDM+3m_nu}, obtained with {\tt CLASS} for two different interaction strengths of $ u_{\nu, \rm DM} \sim 1.65\times10^{-8}$ (red) and $6.68\times 10^{-8}$ (yellow).   Although the onset of the small scale damping looks similar in both the WDM and $\nudm$ scenarios, the presence of oscillations in the latter case may help distinguish them, see e.g.~\cite{Escudero:2018thh, Mosbech:2022uud, Verwohlt:2024efh}.

\subsection{Transfer function}

The difference between the linear matter power spectrum of a NCDM model and that of CDM can be encoded in the transfer function 
\begin{equation}
   T_{i}(k) \equiv \sqrt{\frac{P_{i}(k)}{P_{{\rm CDM}}(k)}}\,,
   \label{eq:transfer_funtion_definition}
\end{equation} 
with $i=$ WDM or $\nudm$ in this paper. 
 In both scenarios, the transfer function can be fit with
\begin{equation}
    f(k, \lcut) \equiv \left\{1+\left(\lcut k
    \right)^\gamma \right\}^{-\delta} \;,
    \label{eq:transf_function_fit_form}
\end{equation}
in rather good
 approximation. The cutoff scale is defined as
\begin{equation}
    \lcut \equiv 
    \begin{cases}
    \displaystyle a_{\rm WDM} \left[\frac{1 ~ {\rm keV}}{m_{\rm WDM}}\right]^{b_{\rm WDM}} ~{\rm Mpc}/h  \quad & {\rm in~the~WDM~scenario} \\[10 pt]
    \displaystyle a_\nudm \left[\frac{u_\nudm}{8.5 \times 10^{-7}}\right]^{b_\nudm}~{\rm Mpc}/h \quad & {\rm in~the~}\nu{\rm DM~scenario},
    \end{cases}
    \label{eq:lambda-cut-definition}
\end{equation}
see e.g.~\cite{Bode:2000gq, Viel:2005qj,  Wilkinson:2013kia}.
Here  we take $\{a_i, b_i,\gamma, \delta\}$ as free parameters.  The functional form of Eq.~(\ref{eq:lambda-cut-definition}) mainly captures the initial drop in the transfer functions. From the expression in Eq.~(\ref{eq:k_free_streaming}) and assuming that $\lambda_{\rm cut}$ scales as $1/k_{\rm fs}$, we expect that $a_{\rm WDM} \sim 0.05$ and $b_{\rm WDM} \simeq 1.2$. Similarly, because the diffusion scale in a coupled fluid is proportional to the square root of the cross-section, we expect that $b_{\nu{\rm DM}} \sim 0.5$. To work in a consistent framework for both scenarios together, as will be necessary for our analysis (see section \ref{sec:analysis}), we simultaneously fit all six parameters to the $T_i(k)$ calculated by \texttt{CLASS} according to the method described in appendix~\ref{app:fit_TF}. The best-fit values are reported in Tab.~\ref{tab:fitting_function_best_fit_values} and are in agreement with the above estimates. Therefore, using this global fit, we establish a one-to-one correspondence between values of $\lambda_{\rm cut}$ and both $m_{\rm WDM}$ and $u_{\nu\rm DM}$.

\begin{table}[t]
    \centering
    \begin{tabular}{|c|c|c|c|c|c|}
        \hline
          & $a_i$ & $b_i$ & $\gamma$ & $\delta$ \\
        \hline
        WDM & $0.013565$ & $1.1701$ & \multirow{2}{*}{$1.9787$} & \multirow{2}{*}{$38.974$} \\
        $\nu$DM &  $0.014115$ & $0.49566$ & & \\
        \hline
    \end{tabular}
    \caption{Best-fit values for the transfer function fit of Eqs.~(\ref{eq:transf_function_fit_form}) and~(\ref{eq:lambda-cut-definition}). The value of the parameters $a_i$ and $b_i$ depends on the model $i=$WDM or $\nudm$ while $\gamma$ and $\delta$ are shared.}
    \label{tab:fitting_function_best_fit_values}
\end{table}

In the left panel of Fig.~\ref{fig:MPS&TF-LCDM+3m_nu}, we show the ${\rm CDM} + 3m_\nu$, WDM and $\nudm$ linear matter power spectra in solid black, dotted red and yellow and dashed red and yellow, respectively. The solid lines show the fitted transfer function for two values of $\lcut$ (or $m_{\rm WDM}$ or $u_\nudm$) of Eq.~(\ref{eq:transf_function_fit_form}) using the best-fit values we obtained. The red and dashed yellow sets of lines refer to two different values of the cutoff scale $\lcut$, which we either translate to $m_{\rm WDM}$ or $u_\nudm$ respectively using Eq.~(\ref{eq:lambda-cut-definition}). In Tab.~\ref{tab:lcut_conversion}, we provide a conversion between three benchmark values of $\lcut$ used in the plots of this paper and the corresponding $m_{\rm WDM}$ and $u_\nudm$. We use those values to compute the matter power spectrum  using \texttt{CLASS} (in the dotted and dashed cases) or plug them into Eq.~(\ref{eq:transf_function_fit_form}) (in the solid case). The right panel of  Fig.~\ref{fig:MPS&TF-LCDM+3m_nu} shows the corresponding transfer functions.

In Section~\ref{sec:Fish}, we will use Eq.~(\ref{eq:lambda-cut-definition}) to ensure consistency in our Fisher matrix sensitivity analysis across both NCDM scenarios. This approach also allows us to assess whether Eq.~(\ref{eq:transf_function_fit_form}) adequately captures the relevant features of the WDM and $\nudm$ matter power spectra. More specifically, we will investigate whether the acoustic damping features present in the $\nudm$ transfer function could leave observable imprints in the $21\,$cm power spectrum as measured by HERA. Before doing so, in the following sections, we describe our treatment of the halo mass function. Simulations of the $21\,$cm signal are performed with a modified version of the public code {\tt 21cmFAST}~\cite{Mesinger:2010ne} called {\tt 21cmCLAST}\footnote{\url{https://github.com/gaetanfacchinetti/21cmCLAST}}   \cite{Facchinetti:2025hou} with an improved coupling to the {\tt CLASS} version including massive neutrino interactions with DM~\cite{Mosbech:2020ahp}\footnote{\url{https://github.com/MarkMos/CLASS_nu-DM}} and new functionalities. In the latter framework, we detail the NCDM impact of the suppression of the matter power spectrum on the $21\,$cm signal.

\begin{table}[t]
    \centering
    \begin{tabular}{|c|c|c|c|}
    \hline
      $\lcut \;  [ {\rm Mpc}/h ] $   & $1\times10^{-3}$ & $2\times10^{-3}$ & $4\times10^{-3}$ \\
        \hline
        $m_{\rm WDM}\;[\rm keV]$ & $9.29$ & $5.13$ & $2.84$ \\
        \hline
        $u_\nudm $ & $4.07\times10^{-9}$ & $1.65\times10^{-8}$ & $6.68\times10^{-8}$\\
        \hline
    \end{tabular}
    \caption{Corresponding $m_{\rm WDM}$ (center) and $u_\nudm$ (bottom) to a selection of values of $\lcut$ (top).}
    \label{tab:lcut_conversion}
\end{table}

\section{Halo mass functions}
\label{sec:HMF}

With the onset of the matter dominated era, under gravitational attraction the initial linear perturbations begin to grow into the nonlinear regime until they collapse to form virialised DM halos. Because DM halos become the hosts of galaxies when star formation begins, their distribution has a strong effect on the $21\,$cm signal.

\subsection{Spherical collapse}

The excursion set theory allows to derive an analytical expression for the halo mass function (HMF) studying the Brownian motion of the   linear density contrast, see e.g. Press-Schechter \cite{Press:1973iz} and Sheth-Tormen \cite{Sheth:1999mn}.  
Let us assume that structures form from spherical collapse, when $\delta^w_R$,  the linear density contrast smoothed over a region of size $R$ using a window function $w$, reaches $\Delta_c(z) = \delta_{\rm c}D(0)/D(z)$, where $D$ is the growth factor and $\delta_{\rm c} = 1.686$. At redshift $z$ inside a region of variance $S_0$ with average linear density contrast $\delta_0 < \Delta_c(z)$, the HMF is given by
\begin{equation}
   \frac{\partial n^w(M, z; \delta_0, S_0)}{\partial \ln M} = \frac{\overline{\rho}_{\rm m, 0}}{\sqrt{2\pi}}\frac{\Delta_c(z) - \delta_0}{[\mathcal{S}^w(M) - S_0]^{3/2}} \exp \left\{ - \frac{\left(\Delta_c(z) - \delta_0\right)^2}{ 2[\mathcal{S}^w(M) - S_0]}\right\} \left| \frac{{\rm d} \mathcal{S}^w}{{\rm d} M} \right|~\,,
   \label{eq:HMF}
\end{equation}
where $\overline{\rho}_{\rm m, 0}$ is the average matter density today and the function $\mathcal{S}^w : M \mapsto S^w(R)$ associates a halo mass to the variance of the linear matter power spectrum within the given region of size $R$. It is defined as
\begin{equation}
    S_R^w \equiv \left< \left(\delta_R^w\right)^2 \right> = \int \frac{{\rm d}^3 {\bf k}}{(2\pi)^3} P(k) |\hat W_R^w(k)|^2 \, ,
    \label{eq:variance}
\end{equation}
where $P(k)$ again denotes the linear matter power spectrum of the scenario under study.
In this paper, we set the window function to a sharp-$k$ ($w=\rm SK$) window function defined as
\begin{equation}
    \hat W_R^{\rm SK}(k) = \Theta(1-kR) \, .
\label{eq:SK}
\end{equation}

Note that it is with such a window function that the analytical form of the HMF was initially derived. We furthermore choose to use the SK window function in this analysis because it has been shown that in the computation of the HMF, it properly accounts for the presence of NCDM \cite{Benson:2012su, Schneider:2013ria}. The top-hat (TH) window function is usually used a posteriori for CDM, as implemented by default in {\tt 21cmFAST v3}, notably because it shows a clear correspondence between the mass of a halo and its radius, $R_{\rm TH}(M)$, with $M= 4/3\pi\overline{\rho}_{\rm m, 0}R_{\rm TH}^3 $. The only drawback of the sharp-$k$ window function is, indeed, that it does not define a finite volume in real space. Nonetheless, a simple prescription can be used with the halo radius $R_{\rm SK}(M)$ to satisfy:
\begin{equation}
 M = \frac43\pi\overline{\rho}_{\rm m, 0}(c_{\rm SK}R_{\rm SK})^3\, .
  \label{eq:RSK}
\end{equation}
An analytic solution for $c_{\rm SK}$ is to define the sharp-$k$ volume $V(R)$ inside a region of radius $R$ such that $W_R^{\rm SK}(0) V(R) = 1$, which gives $V(R) = 6\pi^2 R^3$ and $c_{\rm SK} = (9\pi^2/2)^{1/3} \simeq 2.42$. Another solution is to directly rely on fits obtained from numerical simulations. This is the approach that we use in this work,  using the radius-to-mass conversion factor $c_{\rm SK}=2.5$ obtained in Ref.~\cite{Schneider:2014rda} from numerical simulation assuming the same HMF as the one used in our analysis (described in Sec.~\ref{sec:ellipscoll}). This choice is also in good agreement with the analytical prescription discussed above ($c_{\rm SK}=2.42$).

The importance of the sharp-$k$ window function appears when computing the HMF with a rapidly varying matter power spectrum \cite{Benson:2012su, Schneider:2013ria, Leo:2018odn}. In the case of a sudden depletion on small scales, when $M \to 0$, the variance $\mathcal{S}^w(M)$ becomes a constant and the shape of the HMF is mostly driven by ${\rm d} \mathcal{S}^w / {\rm d} M$. This derivative is highly sensitive to the exact shape of any smooth window function. Adopting a window function different from the sharp-$k$ actually  leads to unphysical HMFs.  This is discussed in more details in App.~\ref{sec:HMFapp}.

\subsection{Implementation with ellipsoidal collapse}
\label{sec:ellipscoll}
In this work, we do not work with Eq.~(\ref{eq:HMF}) directly as we take into account ellipsoidal collapse. This leads to a correction factor in the HMF that takes the form of a normalisation factor to the collapsed fraction (see Ref.~\cite{Mesinger:2010ne} for more details). More precisely, this correction factor is evaluated from the Sheth and Tormen HMF \cite{Sheth:1999mn} which, for a sharp-$k$ window function, takes the form:
\begin{equation}
    \left.  \frac{\partial n^{\rm SK}(M, z)}{\partial \ln M} \right|_{\rm ST}= \frac{\overline{\rho}_{\rm m, 0}}{\sqrt{2\pi}} \frac{\Delta_c(z)}{\left(\mathcal{S}^{\rm SK}(M)\right)^{3/2}}A\sqrt{q}\left(1+\left(\frac{q\Delta_c^2(z)}{\mathcal{S}^{\rm SK}(M)}\right)^{-p}\right) e^{-\frac{q \Delta_c^2(z)}{2\mathcal{S}^{\rm SK}(M)}}\left| \frac{{\rm d} \mathcal{S}^{\rm SK}}{{\rm d} M} \right| \, .
\end{equation}
In this expression, $A = 0.322$, $p = 0.3$ and $q=1$.

\begin{figure}[t]
\begin{center}
\includegraphics[width=0.75\linewidth]{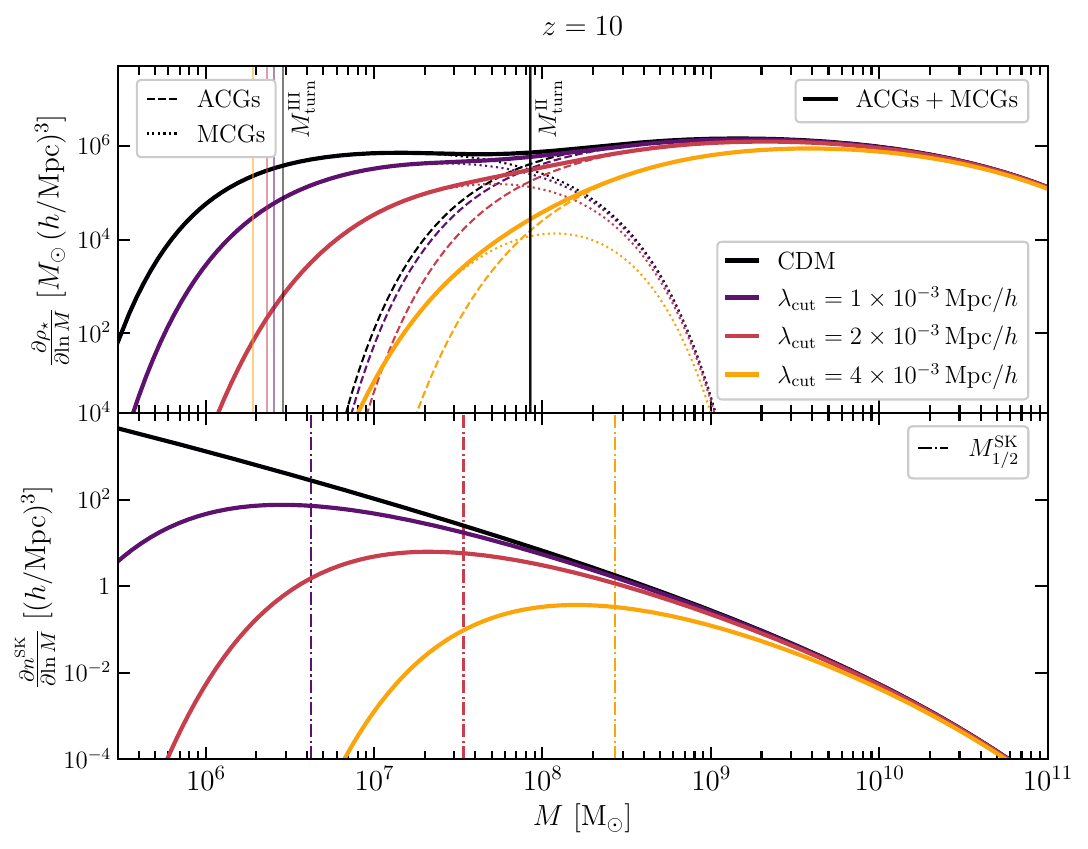}
\caption{ 
  {\bf Upper panel} Stellar mass distribution in ACGs and MCGs   assuming CDM (black) and for three cutoff scales (coloured) 
  considering the reference astrophysical model of Tab.~\ref{tab:fiducial_astro_parameters} and using the fitting function of Eq.~(\ref{eq:transf_function_fit_form}) to obtain the NCDM matter power spectrum. The different line styles show the respective contributions from ACGs and MCGs. The vertical lines in the upper panel indicate the values of $M_{\rm turn}^{\rm II}$ and $M_{\rm turn}^{\rm III}$. {\bf Lower panel} Corresponding halo mass function for CDM and the three values of $\lambda_{\rm cut }$ considered in the upper panel.  The dash-dotted vertical lines show the halo mass $ M^{\rm SK}_{1/2}$ corresponding to a wave number at which the transfer function is equal to $1/2$, see text for details.}

\label{fig:stellar_density_FR}
\end{center}
\end{figure}

In the bottom panel of Fig.~\ref{fig:stellar_density_FR}, we show the resulting HMF for CDM (black) and for three non-zero values of the cutoff scale (purple to yellow colours for increasing $\lcut$ or equivalently warmer or more strongly interacting DM). The HMF in Fig.~\ref{fig:stellar_density_FR} is computed assuming the fitted transfer function of Eq.~(\ref{eq:transf_function_fit_form}). The dash-dotted vertical lines in both figures indicates the halo mass $M^{\rm SK}_{1/2}$, which corresponds to a halo of radius $R^{\rm SK}=k_{1/2}^{-1}$ in Eq.~(\ref{eq:RSK}), where $k_{1/2}$ is the wavemode for which the transfer function is equal to $1/2$. Using the best-fit values from Tab.~\ref{tab:fitting_function_best_fit_values} yields $k_{1/2}=0.13/\lcut$. It is visible that the abundance of halos with masses $\lesssim M^{\rm SK}_{1/2}$ becomes significantly suppressed. For completeness, the HMF equivalent to the ones of Fig.~\ref{fig:stellar_density_FR} obtained assuming WDM or $\nudm$ cosmology  are shown in the lower panel of Fig.~\ref{fig:hmf_model_comparison}. 

\section{NCDM imprint on the $21\,$cm signal}
\label{sec:21cm}

In this section we summarise the basics of the $21\,$cm cosmology and introduce the $21\,$cm power spectrum that will be the observable of HERA. Then, we discuss the astrophysical modelling of star formation as implemented in {\tt version 3}  of {\tt 21cmFAST} and the degeneracies between the astrophysical parameters and the WDM or $\nudm$ parameters.

\subsection{The $21\,$cm signal in cosmology}

As the dominant constituent of baryonic matter, hydrogen plays a key role in structure formation. Tracing its distribution and evolution during cosmic dawn is essential to probe the astrophysical and cosmological processes that shaped the early universe. A useful method for this relies on spin-flip transitions between the two ground state energy levels  of neutral hydrogen that give rise to absorption or emission of photons of wavelength $\sim 21\,\textrm{cm}$. The relative occupancy of the two ground state energy levels is quantified by the spin temperature $T_{\rm S}$. Processes that affect the spin temperature include the emission and absorption of CMB photons, scatterings on other atoms and resonant Lyman-$\alpha$ transitions. For the redshift range of interest, it is given by its equilibrium value that  can be expressed as~\cite{Pritchard:2011xb}
\begin{align}
    T_{\rm S}^{-1}=\frac{T_{\textrm{CMB}}^{-1}+(x_{\alpha}+x_{\rm c})T_{\rm K}^{-1}}{1+x_{\alpha}+x_{\rm c}}\,,
\end{align}
where $T_{\rm K}$ is the kinetic temperature of the IGM, $x_{\rm c}$ encodes the efficiency of collisions within the gas and $x_\alpha$ the coupling to Lyman-$\alpha$ photons, see also~\cite{Flitter:2024eay}. In this expression we have made the assumption that the colour temperature is equal to the kinetic temperature $T_{\rm c} = T_{\rm K}$ \cite{Hirata:2005mz}.

When the  redshifted $21\,$cm signal is observed in contrast with the CMB background it can be expressed in terms of the differential brightness temperature as
\begin{equation}
    \delta T_{\rm 21} \equiv \frac{(T_{\rm S}-T_{{\rm CMB}})}{(1+z)}\left(1-e^{-\tau_\nu}\right)\,. 
    \label{eq:Def_of_TempBrightnessVar}
\end{equation}
In this expression, $\tau_{\nu}$ is the optical depth to $21\,$cm photons and $\delta T_{21}\equiv \delta T_{21}(\textbf{x},z)$ depends on the line of sight. To a  good approximation, the differential brightness temperature can be expressed as \cite{Furlanetto:2006jb}
\begin{align}
    \delta T_{\rm 21}\approx 20\,\textrm{mK}\left(\frac{T_{\rm S}-T_{\textrm{CMB}}}{T_{\rm S}}\right)x_{\textrm{HI}}(1+\delta_{\rm b})\left(1+\frac{1}{H}\frac{\textrm{d}v_r}{\textrm{d}r}\right)^{-1}\sqrt{\frac{1+z}{10}\frac{0.15}{\Omega_{\rm m} h^2}}\frac{\Omega_{\rm b} h^2}{0.023}\,,
\end{align}
where $x_{\textrm{HI}}$ is the neutral fraction of hydrogen, $\delta_{\rm b}\equiv\delta_{\rm b}(\textbf{x},z)$ is the baryon density contrast and $\textrm{d}v_r / \textrm{d}r$ is the velocity gradient along the line of sight.
For more comprehensive information, see~\cite{Furlanetto:2006jb,Pritchard:2011xb}.
In practice, current and next-generation telescope arrays like HERA will measure  the quantity $\overline{\delta T_{\rm 21}}^2(z)\Delta^2_{21}(k,z)$, where $\Delta^2_{21}$ denotes  the dimensionless power spectrum of perturbations in the $21\,$cm signal
\begin{equation}
        \Delta_{21}^2(k, z) = \frac{2k^2}{\pi} \int_0^{\infty} \xi_{\rm 21}(r, z) \sin(kr)r {\rm d} r \, ,
\end{equation}
where $\xi_{\rm 21}$ is the two-points correlation function of the brightness temperature contrast,
\begin{equation}
    \xi_{\rm 21}(|{\bf x} - {\bf y}|, z) \equiv \left<  \delta_{\rm 21}({\bf x}, z)  \delta_{\rm 21}({\bf y}, z) \right> \quad {\rm with} \quad   \delta_{\rm 21}({\bf x}, z) \equiv \frac{\delta T_{\rm 21}({\bf x}, z)}{\overline{\delta T_{\rm 21}}(z)}-1 \, .
\end{equation}
The brackets $\left< \dots \right>$ refer to the ensemble average while the bar in $\overline{\delta T_{\rm 21}}(z)$ refers to the spatial average. In our pipeline, the $21\,$cm power spectra are evaluated from the output of {\tt 21cmFAST} using the python package {\tt powerbox} \cite{Murray2018}. In the following we will always consider the quantity $\overline{\delta T_{\rm 21}}^2(z)\Delta^2_{21}(k,z)$ as the $21\,$cm observable. The details of the model implemented in {\tt 21cmFAST} are given in the next subsection.

\vspace*{12pt}

\subsection{Star formation, ionisation and X-ray emission}
\label{sec:SFR-ion-Xrays}

We model the astrophysics using the implementation of {\tt
  21cmFAST\,v3}, as described in Ref.~\cite{Park:2018ljd}, that includes
a prescription for both atomic  and molecular
cooling galaxies. The latter are expected to be hosted by
minihalos. The stellar mass in a halo of virial mass $M$  is parametrised by
\begin{equation}
      M_\star^i(M) \equiv  M f_\star^{i}(M) \,,
\end{equation} 
where $i = {\rm II, III}$ correspond to PopII or PopIII stars, assumed to be hosted by ACG and MCG respectively, see e.g.~Ref.~\cite{Munoz:2021psm}. $f_\star^i$ is the stellar-to-halo mass ratio. Both $f_\star^i$ and the fraction of ionising ultraviolet (UV) photons that escape the galaxies, $f_{\rm esc}$, are parametrised with a power law~\cite{Qin:2020xyh, Stefanon_2021, Shuntov:2022qwu}
\begin{equation}
\begin{split}
    f_\star^i(M) & =  {\rm min}\left\{ 1, f_{\star, \ell_i}^i \left(\frac{M}{10^{\ell_i} ~{\rm M_\odot}}\right)^{\alpha_\star^i} \right\} \frac{\Omega_{\rm b}}{\Omega_{\rm m}} \\
        f_{\rm esc}^i(M) & =  {\rm min}\left\{ 1, f_{{\rm esc}, \ell_i}^i \left(\frac{M}{10^{\ell_i} ~{\rm M_\odot}}\right)^{\alpha_{\rm esc}} \right\} \, ,
    \end{split}
    \label{eq:f_star_f_eq}
\end{equation}
where $\ell_i$ are set to $\ell_{\rm II} = 10$ and $\ell_{\rm III} = 7$ respectively -- to match with the characteristic mass scales of ACG and MCG hosting halos. Here,  $f_{\star, \ell_i}^{i}$, $\alpha_\star^i$, $f_{{\rm esc}, \ell_i}^i$ and $\alpha_{\rm esc}$ are $7$ new free parameters.  The  inefficiency of star formation in low-mass halos is captured by a duty cycle parameter, conventionally expressed as~\cite{Qin:2020xyh,Munoz:2021psm}
\begin{equation}
\begin{split}
    f_{\rm duty}^{\rm II}(M, z) & = \exp\left(-\frac{M_{\rm turn}^{\rm II}(z)}{M}\right) \\ f_{\rm duty}^{\rm III}(M, z) & = \exp\left(-\frac{M_{\rm turn}^{\rm III}(z)}{M}\right) \exp\left(-\frac{M}{M_{\rm atom}(z)}\right)
    \end{split}
\end{equation}
for ACGs and MCGs respectively where $M$ denotes the halo mass. The second expression assumes that there is a smooth
transition from MCGs,  with halo mass of  $M<M_{\rm atom}$ that form PopIII stars, to
ACGs, with halo mass $M>M_{\rm atom}$ that form PopII stars. The atomic cooling mass threshold is given by $M_{\rm atom} = 5 \times 10^7 \, [(1+z)/10]^{-3/2} ~{\rm M_\odot}$~\cite{Barkana:2000fd}.
For the ACGs, $M_{\rm turn}^{\rm II} = {\rm max}(M_{\rm atom}, M^{\rm RE}_{\rm crit})$ where $M^{\rm RE}_{\rm crit}$ is a characteristic mass below which photoheating during reionisation is able to significantly suppress gas content \cite{Sobacchi:2013ww}. For the MCGs, $M_{\rm turn}^{\rm III} = {\rm max}(M_{\rm mol}, M^{\rm RE}_{\rm crit})$ where $M_{\rm mol}$ is the molecular cooling mass threshold that accounts for the photodissociation of ${\rm H}_{2}$ molecules from photons in the Lyman-Werner band ($11.2$--$13.6$ eV) \cite{Munoz:2019hjh}.

The amounts of UV-ionising, Lyman-$\alpha$ or X-ray photons injected in the IGM by the galaxies  depend on the stellar mass distribution, defined as the product of the halo mass function, the duty cycle and the stellar mass,
\begin{equation}
  \frac{\partial \rho_\star^i(M, z; \delta_R, S_R)}{\partial \ln M} \equiv \frac{\partial n^{\rm SK}(M, z; \delta_R, S_R)}{\partial \ln M} f_{\rm duty}^i(M) M_\star^i(M, z) \, .
  \label{eq:SMD}
\end{equation}
The X-ray specific emissivity at position ${\bf x}$ and redshift $z$ with energy $E$ is implemented as
\begin{equation}
    \epsilon_X({\bf x}, E, z) = [1+\overline {\delta_{\rm b}}({\bf x}, z)] \frac{ \mathcal{L}^i_X E^{-\alpha_X}}{t_\star H^{-1}(z)}\sum_{i \in \rm \{II, III\}} \int_0^\infty \frac{\partial \rho_\star^i(M, z; \delta_R, S_R)}{\partial \ln M}  {\rm d} \ln  M\,,
    \label{eq:epsX}
\end{equation}
where $t_\star$ is a free parameter between zero and one that sets the star formation rate with respect to the Hubble rate, $\overline{\delta_{\rm b}}({\bf x}, z)$ is the average value of the nonlinear matter density contrast of the shell around $({\bf x}, z)$, and $\mathcal{L}_X^i$ is the normalisation of the X-ray luminosity per star formation rate (SFR) \cite{Das:2017fys}. In practice, $\mathcal{L}_X^i$ is not used as a parameter of the model but rather $L^i_X$, defined as the integrated soft-band ($E_0 < E < 2~{\rm keV}$) luminosity per SFR (in units of $\rm erg ~yr~s^{-1}~M_\odot^{-1}$) according to
\begin{equation}
   L^i_X =  \mathcal{L}^i_X \int_{E_0}^{2~{\rm keV}} {\rm d } E E^{-\alpha_X}\,.
\end{equation}
Here, $E_0$  is the threshold energy for X-rays to escape the host galaxy (and thus heat the IGM). Similarly, the amount of injected ionising UV photons is
\begin{equation}
    n_{\rm ion}({\bf x}, z) = \frac{N_{\gamma / {\rm b}}}{\rho_{\rm b}(z)} \sum_{i \in \rm \{II, III\}}  \int_0^\infty   \frac{\partial \rho_\star^i(M, z; \delta_R, S_R)}{\partial \ln M} f_{\rm esc}^i(M)  {\rm d} \ln M\,
    \label{eq:nion}
\end{equation}
where $N_{\gamma / {\rm b}}$ is the number of ionising photons per stellar baryon fixed to 5000 (otherwise largely degenerate with $f_{\star, \ell_i}^i$). Apart of the introduction of the escape fraction, both integrals in equations~(\ref{eq:epsX}) and (\ref{eq:nion}) are similar and mainly depend on the stellar mass distribution into ACGs and MCGs.

\begin{table}[t]
    \centering
    \begin{tabular}{|c c c c c c c|}
        \hline
        $\log_{10}f_\star^{\rm II}$ &  $\alpha_\star^{\rm II}$ & $t_\star$ & $\log_{10}f_{\rm esc}^{\rm II}$ & $\alpha_{\rm esc}$ & $\displaystyle\log_{10}\frac{L_X^{\rm II}}{{\rm erg\,s}^{-1}\,{\rm M}_\odot^{-1}\,\rm yr}$ & $E_0\;/\,\rm eV$ \\
        \hline
        $-1.25$ & $0.5$ & $0.5$ & $-1.35$ & $-0.3$ & $40.5$ & $500$ \\
        \hline
        \hline
        $\log_{10}f_\star^{\rm III}$ & $\alpha_\star^{\rm III}$ & & $\log_{10}f_{\rm esc}^{\rm III}$ & & $\displaystyle\log_{10}\frac{L_X^{\rm III}}{{\rm erg\,s}^{-1}\,{\rm M}_\odot^{-1}\,\rm yr}$ & \\
        \hline
        $-2.5$ & $0.1$ & & $-1.35$ & & $40.5$ & \\
        \hline
    \end{tabular}

    \caption{Reference values of the astrophysical parameters used throughout this work. They are based on the fiducial values from the EOS2021 model introduced in~\cite{Munoz:2021psm}, albeit with $\alpha_\star^{\rm III}$ changed from $0.0$ to $0.1$, for the purpose to ease the Fisher forecast treatment. In the upper panel we report the ACG and common parameter values while in the lower panel we report the corresponding MCG ones. }
    \label{tab:fiducial_astro_parameters}
\end{table}

In Tab.~\ref{tab:fiducial_astro_parameters}, we summarise the value of the astrophysical parameters that we consider in this work. We use the EOS2021 model of~Ref.~\cite{Munoz:2021psm} as our reference model, albeit with $\alpha_\star^{\rm II}$ changed from 0 to 0.1 to simplify the computation of derivatives in the Fisher analyses. This model is compatible with the current best constraints on the $21\,$cm power spectrum \cite{HERA:2021noe} and, according to the authors of Ref.~\cite{Munoz:2021psm}, it is in agreement with the optical depth to reionisation constraint from Planck \cite{Planck:2018vyg}. 
The resulting stellar mass distribution of Eq.~(\ref{eq:SMD}) at $z=10$ is shown as a function of the halo mass in the top panel of Fig.~\ref{fig:stellar_density_FR}  for CDM (black) and for three values of the NCDM cutoff scale. For the NCDM curves, we use the fitted transfer function of Eq.~(\ref{eq:transf_function_fit_form}) as input.  At large halo masses, ACGs dominate (dotted lines), while MCGs contribute mainly at smaller masses (dashed lines), producing two distinct bumps. The imprint of the NCDM cutoff scales is discussed in the next section.

\subsection{Dark matter and  astrophysics imprint}
\label{sec:effect}

The effect of non-cold dark matter is expected to be at least partially degenerate with other astrophysical parameters, see e.g.~\cite{Giri:2022nxq,
Hibbard:2022sng, Dey:2022ini,
Schosser:2024aic,Decant:2024bpg}. More specifically, we expect a change in the previously defined cutoff scale to at least partially mimic the effects of changes in the astrophysical parameters governing the star formation rate and efficiency. In this section, we illustrate this with the parameters $f_\star^{\rm III}$ and $\alpha_\star^{\rm II}$.  Furthermore, large cutoff scales might prevent us from efficiently probing MCG parameters, because the abundance of the minihalos in which they form would be too suppressed, see also~\cite{Giri:2022nxq,Decant:2024bpg}. This is already well-visible in the top panel of Fig.~\ref{fig:stellar_density_FR}, which illustrates the impact of different cutoff scales (coloured lines) on the stellar mass distribution introduced in Eq.~(\ref{eq:SMD}) with respect to the CDM scenario (black). It can be seen that increasing values of the cutoff scale $\lcut$ (purple to yellow colouring), equivalent to warmer or more strongly interacting DM scenarios, lead to a progressively stronger suppression of the MCG contribution (dotted lines). The CDM case (black) is recovered in the limit $\lcut\to0$. In the yellow case ($\lcut=4\times10^{-3}\,{\rm Mpc}/h$), the MCG contribution is so suppressed that we will not be able to reconstruct well the associated model parameters. In this case, we expect $\lcut$ to mainly be degenerate with ACG parameters, while for lower values of the cutoff scale we expect stronger degeneracies with MCG parameters to appear.

\begin{figure}[t]
    \centering
    \includegraphics[width=0.6\linewidth]{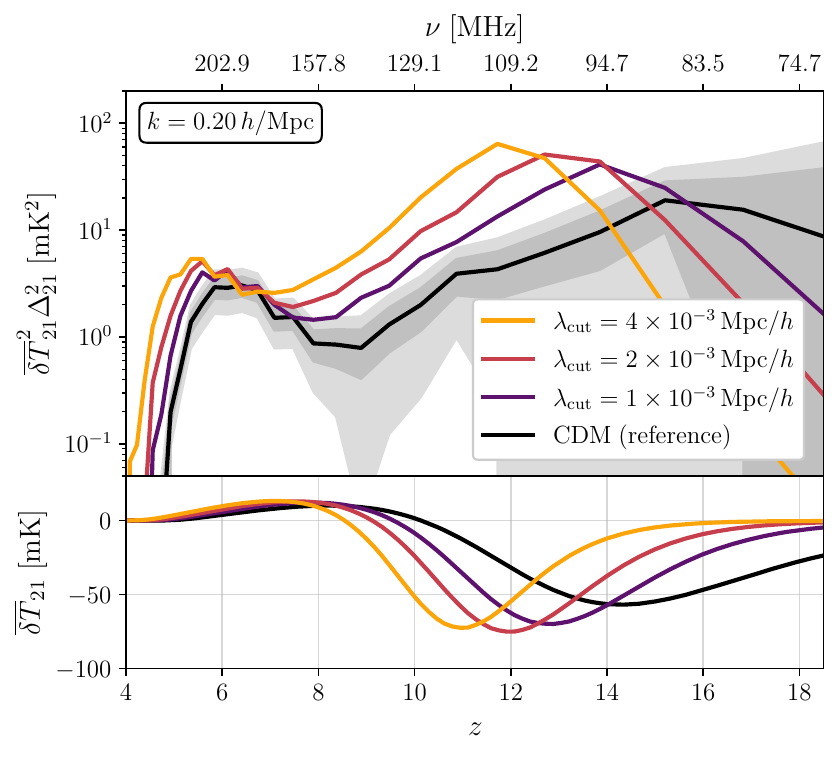}
    \caption{NCDM imprint on the $21\,$cm signal as a function of  the redshift using the fitting function as an input. The upper panel displays the power spectrum at wavelength $k=0.20\;h/\rm Mpc$, the lower panel shows the global signal. The black line assumes CDM and the coloured lines show the effect of different values of $\lcut$. The dark and light shaded bands respectively indicate the associated 1 and 2$\sigma$ measurement noise of HERA, assuming 1000 hours of observation and a 20\% modelling error (see also section~\ref{sec:analysis}). We assume our reference astrophysical model.}
    \label{fig:astro-impact-PS-1}
\end{figure}

Also note that Fig.~\ref{fig:stellar_density_FR} was made using the fitted transfer function of Eq.~(\ref{eq:transf_function_fit_form}). The slightly different linear matter power spectra in the WDM and $\nudm$ scenarios lead to different halo mass functions, which is illustrated for two values of $\lcut=\{1\,{\rm and\;}2\}\times10^{-3}\,{\rm Mpc}/h$ in the appendix in lower panel of Fig.~\ref{fig:hmf_model_comparison}. More concretely, the additional power on small scales in the linear matter power spectrum in the $\nudm$ scenario (compare Fig.~\ref{fig:MPS&TF-LCDM+3m_nu}) leads to an increased abundance of small halos compared to the WDM scenario, as visible in the upper panel of Fig.~\ref{fig:hmf_model_comparison}. This in turn leads to an increased MCG stellar mass distribution, although this effect is subdominant w.r.t. the overall suppression of the MCG contribution induced by the NCDM.

\begin{figure}[t]
    \centering
       \includegraphics[width=0.49\linewidth]{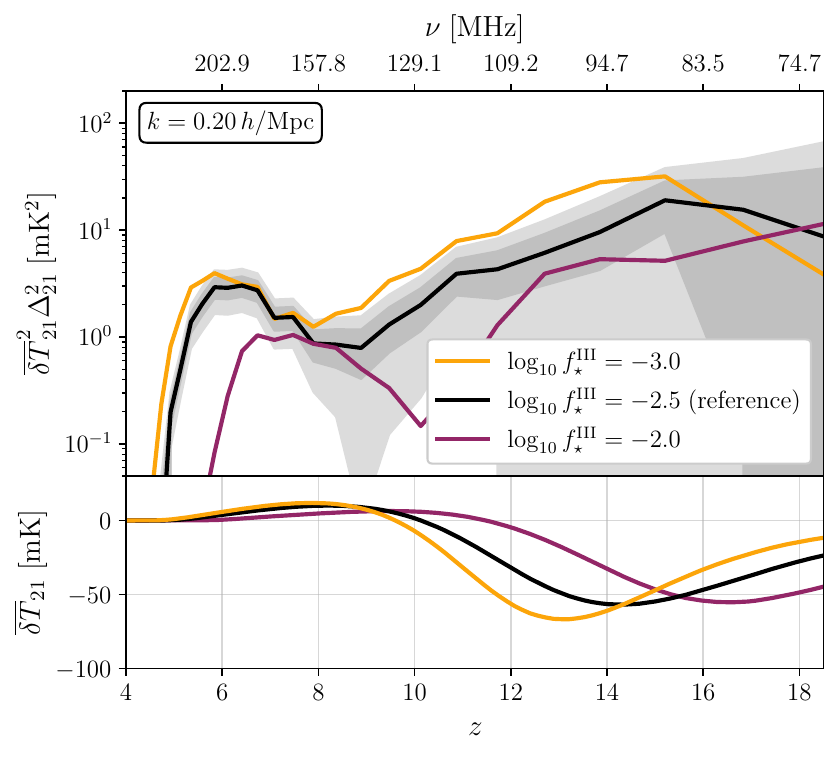}
     \includegraphics[width=0.49\linewidth]{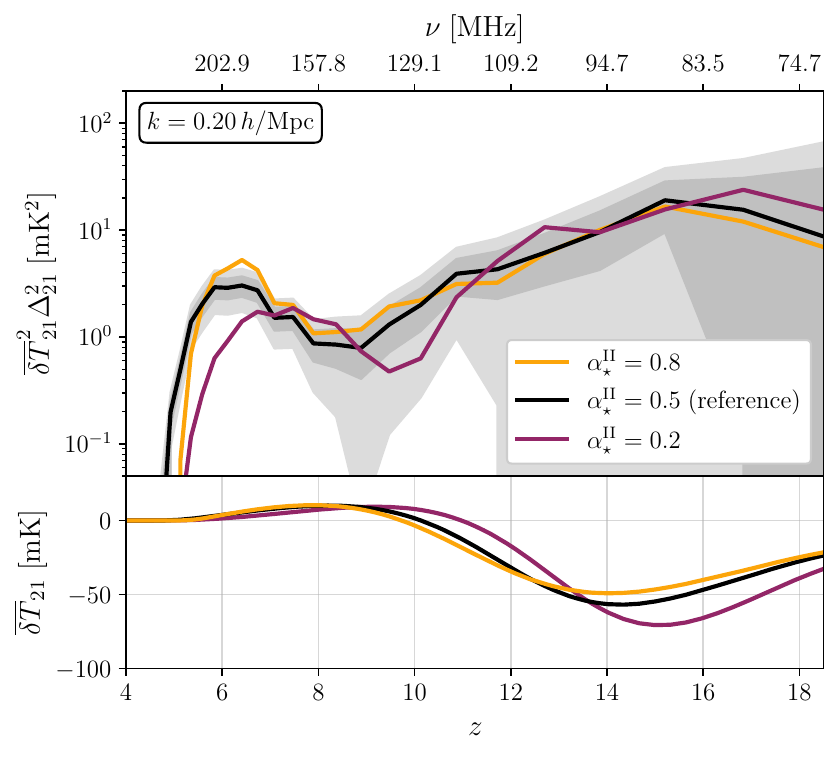}
    \caption{Same as Fig.~\ref{fig:astro-impact-PS-1}, but assuming CDM and showing the effect of varying $\alpha_\star^{\rm III}$ (left panel) and  $\alpha_{\rm esc}$ (right panel). }
    \label{fig:astro-impact-PS-2}
\end{figure}

In Figs.~\ref{fig:astro-impact-PS-1} and~\ref{fig:astro-impact-PS-2}, the evolution of the   $21\,$cm power spectrum at $k = 0.20\;h/\mathrm{Mpc}$ (upper panel) and of the sky averaged brightness temperature (lower panel) is shown as a function of redshift. The light and dark gray shaded bands show the experimental uncertainty at 68\% and 95\% CL for the reference model (with no NCDM cutoff) under the assumption of 20\% modelling error, see Sec.~\ref{sec:Fish} for more details. Each plot explores different values of some of the model parameters that are expected to show degeneracies between each other.

Figure~\ref{fig:astro-impact-PS-1} illustrates that an increase in $\lcut$, or equivalently an increase in $u_\nudm$ or a decrease in $m_{\rm WDM}$, effectively delays the evolution of the differential brightness temperature and the $21\,$cm power spectrum, with each of their  features appearing at later times. This happens because larger cut-offs suppresses the matter power spectrum on larger scales, which reduces the abundance of low-mass halos (as visible in the lower panel of Fig.~\ref{fig:stellar_density_FR}). As a consequence, the production of Lyman-$\alpha$, X-ray and UV photons that drive these features is diminished and postponed.

A reduced stellar-to-halo mass ratio in MCGs, meaning a lower $f_{\star}^{\rm III}$, results in a decreased energy injection into the IGM from the lowest mass halos and similarly causes a delay in the features in the power spectrum, as is visible in the left panel of Fig.~\ref{fig:astro-impact-PS-2}. Conversely, a large positive value of $\alpha_\star^{\rm II}$, which controls the dependence of the stellar-to-halo mass ratio on the halo mass in Eq.~(\ref{eq:f_star_f_eq}), would analogously decrease the relative contribution of low-mass halos to the signal by suppressing the relative contribution to the Lyman-$\alpha$, X-rays and ionising flux of low-mass halos. Increasing  $\alpha_\star^{\rm II}$ would also shift the $21\,$cm signal towards later times, as shown in the right panel of Fig.~\ref{fig:astro-impact-PS-2}, thereby also mimicking the effect of an NCDM cutoff scale, albeit less significantly. Further discussion of the degeneracies between astrophysical parameters and the NCDM cutoff scale is provided in Sec.~\ref{sec:degen-results} in the context of the analysis our results.

\section{Analysis}
\label{sec:analysis}

We estimate the sensitivity of the completed HERA telescope to both the WDM and $\nudm$ scenarios as well as a scenario using the fitted transfer function, assuming 1000 hours of observation. To that end, we perform a Fisher matrix forecast using the same tools as in Ref.~\cite{Facchinetti:2023slb}.
In the following, we detail our application of the Fisher matrix formalism to this work and discuss our results.

\subsection{Fisher matrices}
\label{sec:Fish}

The Fisher information matrix for a set of parameters $\{\theta_i\}$ is defined as
\begin{align}
    F_{ij} \equiv -\mathbb{E}_{\mathcal{O}} \left[ \left. \frac{\partial^2 }{\partial \theta_i \partial \theta_j} \ln \mathcal{L}(\mathcal{O} \, |\, \theta)  \, \right| \, \theta \right]  \, ,
\end{align}
where $\mathcal{L}$ is the likelihood of the observed data given a model and $\mathbb{E}_{\mathcal{O}}$ the associated expectation value.  The minimum covariance of two parameters can then be estimated as $\mathrm{cov}(\theta_i, \theta_j)=(F^{-1})_{ij}$ at the Cramer-Rao bound \cite{frechet1943extension, darmois1945limites, aitken1942xv}, and thus the minimum marginalised standard deviation on a single parameter $\theta_i$ as $\sigma_i= [(F^{-1})_{ii}]^{1/2}$. Assuming a Gaussian Likelihood, we compute Fisher matrix elements as detailed in Ref.~\cite{Facchinetti:2023slb} as
\begin{equation}
  F_{ij}=\sum_{i_k}\sum_{i_z}\frac{1}{\sigma_{21}^2(k_{i_k},z_{i_z})}\frac{\partial {\cal O}(k_{i_k},z_{i_z})}{\partial \theta_i} \frac{\partial {\cal O}(k_{i_k},z_{i_z})}{\partial \theta_j}\,.
\end{equation}
The sums run over all the wavenumber $k$ and redshift $z$ bins, the observable in this case is ${\cal O}=\overline{\delta T}^2_{21} \Delta_{21}^2$ and for the measurement error we assume:
\begin{equation}
    \sigma_{21}^2(i_k,i_z) = \sigma_{\rm exp}^2(i_k,i_z) + \sigma_{\rm shot}^2(i_k,i_z) + [\varepsilon \times {\cal O}(i_k,i_z)]^2\,.
    \label{eq:21cm_noise}
\end{equation}
The first term accounts for the experimental error (i.e. thermal noise) and is evaluated using the public Python package \texttt{21cmSense}\footnote{\url{https://github.com/jpober/21cmSense}}~\cite{Pober:2012zz, Pober:2013jna}, the second term accounts for shot noise in each bin and the third term accounts for the modelling error on $\mathcal{O}$. For the latter, a common practice in the field is to use $\varepsilon=0.2$~\cite{Park:2018ljd}. However, we will find that the reconstruction error and detection threshold that we derive for the cutoff scale will depend strongly on the choice of $\varepsilon$. We will thus also consider the case where $\varepsilon=0$, i.e. no modelling error.

Using the Fisher matrix formalism, the sensitivity of HERA to our NCDM scenarios can be evaluated in several ways.  One approach would be  to estimate bounds on the WDM particle mass and $\nu$DM interaction strength from the $21\,$cm power spectrum, assuming CDM as the fiducial model, as done in e.g.~Ref.~\cite{Facchinetti:2023slb} for decaying DM. However, for non-cold dark matter, this method is hindered by the \emph{locality} of the Fisher matrix formalism, which relies on linear extrapolations around the fiducial point. From the discussion in Sec.~\ref{sec:effect}, we know that the NCDM suppression of low mass halos can mimic the effect of the threshold for star forming galaxies, particularly the one of minihalos set by $M_{\rm turn}^{\rm III}$. Yet, once the NCDM cutoff scale is small enough to significantly suppress the formation of halos of mass $M < M_{\rm turn}^{\rm III}$,  the $21\,$cm power spectrum becomes largely insensitive to variations in $\lcut$. The  effect becomes indeed screened by $M_{\rm turn}^{\rm III}$. As a result, taking CDM as the fiducial model (or equivalently $\lcut\to 0$) would  lead to small corresponding Fisher matrix elements, resulting in large forecasted uncertainties and artificially weak constraints. 

In this work instead, we compute the Fisher matrix elements for a series of fixed NCDM parameters ($\lambda_{\rm cut}$, $m_{\rm WDM}$ or $u_{\nu \rm DM}$) and determine for which fiducial value of those parameters we can claim a detection with a certain confidence level. As a result, we report a detection threshold and not a constraint. Nonetheless, specific attention is required to derive meaningful detection thresholds that can be  compared between the two different NCDM models. Indeed, the posterior distribution for WDM mass and $\nudm$ interaction strength are certainly not both Gaussian (and most likely none of them really is). Therefore, we cannot simply perform two independent Fisher analysis for the two scenarios and use the Fisher estimator of the standard deviation to obtain 95\% CL intervals that can be compared together. The different shapes of the posterior can strongly bias the result. This issue is discussed in details in appendix~\ref{app:fisher_change_variable}, and, in a different context, in e.g.~\cite{Foster:2017hbq, Cowan:2010js}.

To enable direct comparison between WDM and $\nudm$ scenarios, we adopt $\lcut$ as the universal input parameter for all simulations, that is converted to either $m_{\rm WDM}$ or $u_\nudm$ using the fit of Eq.~(\ref{eq:lambda-cut-definition}) when relevant. The resulting quantity is then used as input for \texttt{CLASS} to compute the transfer function $T_i(k)$, which is passed to {\tt 21cmFAST}. More precisely, this conversion and communication pipeline is automated through new functions and interpolation tables implemented in the wrapper of {\tt 21cmCLAST}. The input dictionary for {\tt CLASS} is generated directly from the {\tt 21cmCLAST} input object; {\tt CLASS} is then called via its {\tt python} wrapper, and the resulting transfer function is reformatted and passed to the {\tt C} code for computing the evolution of the IGM. As a result, the user simply needs to specify the desired value of $\lcut$ in the standard {\tt CosmoParams} object and select one of the following methods:
\begin{enumerate}[noitemsep, nolistsep]
\item\label{it:fit} use $\lcut$ directly in the fitted transfer function expression of Eq.~(\ref{eq:transf_function_fit_form}), which is then applied to the CDM matter power spectrum as computed by {\tt CLASS},
\item\label{it:WDM} convert $\lcut$ to a WDM mass, which is then input into {\tt CLASS} to obtain the WDM matter power spectrum,
\item\label{it:nuDM} convert $\lcut$ to a $\nudm$ interaction strength, which is then input into {\tt CLASS} to obtain the $\nudm$ matter power spectrum.
  
\end{enumerate}

\subsection{Confidence intevals}

In all three scenarios,~\ref{it:fit}-\ref{it:nuDM}, $\lcut$ remains the input parameter of the overall call. 
To derive confidence intervals, we assume the Cramér-Rao bound is saturated, so that the Fisher matrix is equivalent to the inverse of a covariance matrix. Under this assumption, one can directly associate a multivariate Gaussian distribution with the parameters -- details of the numerical evaluation of the Fisher matrix are given in appendix~\ref{app:fisher}.
However, we also need to take into account that $\lcut$ cannot be negative, i.e. enforce a (flat) prior $\lcut\in[0,\infty)$ while keeping the posterior normalised. We thus take the marginalised posterior of $\lcut$ to be given by a renormalised, truncated Gaussian probability distribution function
\begin{equation}
    p_{\lcut}(\lcut) = \sqrt{\frac{2}{\pi \sigma^2}} \frac{e^{-\displaystyle\frac{\left(\lcut - \lcutfid\right)^2}{2\sigma^2}}}{1+{\rm erf}\left(\displaystyle \frac{\lcutfid}{\sqrt{2}\sigma}\right)} \;\,{\rm for}\; \lcut\geq0
    \label{eq:pdflcut}
\end{equation}
and $0$ otherwise, where $\sigma^2=(F^{-1})_{\lcut\lcut}$. Note that the variance ${\rm Var}(\lcut)$ of this distribution $p_{\lcut}(\lcut)$ is not given by $\sigma^2$ unless $\lcutfid\to\infty$.

We can now introduce a confidence interval at confidence level $\alpha$ by solving for $\delta_\lambda$ in
\begin{equation}
    \int_{\max(\lcutfid-\delta_\lambda,\;0)}^{\lcutfid+ \delta_\lambda} p_{\lcut}(x)  {\rm d} x = \alpha.
\end{equation}
If we find that $\delta_\lambda < \lcutfid$, then the confidence interval is 
$[\lcutfid - \delta_\lambda, \lcutfid + \delta_\lambda]$ and is thus symmetric with respect to the fiducial value. This corresponds to a detection of $\lcut$ (or the associate DM mass or interaction strength depending on the chosen model) at confidence level $\alpha$. On the contrary, if $\delta_\lambda \ge \lcutfid$, the confidence interval is $[0, \lcutfid + \delta_\lambda]$ and we only obtain an upper limit on $\lcut$.

\section{Results}
\label{sec:results}

In this section we present the results of our Fisher matrix analysis, performed as described in Sec.~\ref{sec:analysis} under the assumption of the astrophysical model parameters given in Tab.~\ref{tab:fiducial_astro_parameters} and the fully built HERA telescope with 1000 hours of observation. We first report on the forecast for the detection threshold of $\lcut$ in Sec.~\ref{sec:threshold}. In particular, the marginalised posteriors of the reconstructed $\lcut$ normalised to their fiducial values are shown in Fig.~\ref{fig:detection_threshold_plot}. We further comment on degeneracies between astrophysical and NCDM parameters in Sec.~\ref{sec:degen-results} with the two-dimensional posteriors of a selection of parameters shown in Fig.~\ref{fig:wdm_triangle_plot_small}. Finally, we discuss the difficulty to distinguish between the WDM and $\nudm$ models in Sec.~\ref{sec:disctinction} and provide a $\Delta \chi^2$ estimate of the capacity of a "test" WDM scenario to account for a "true" $\nudm$ scenario.

\subsection{Detection threshold}
\label{sec:threshold}

\begin{figure}[t]
    \centering
    \includegraphics[width=0.99\linewidth]{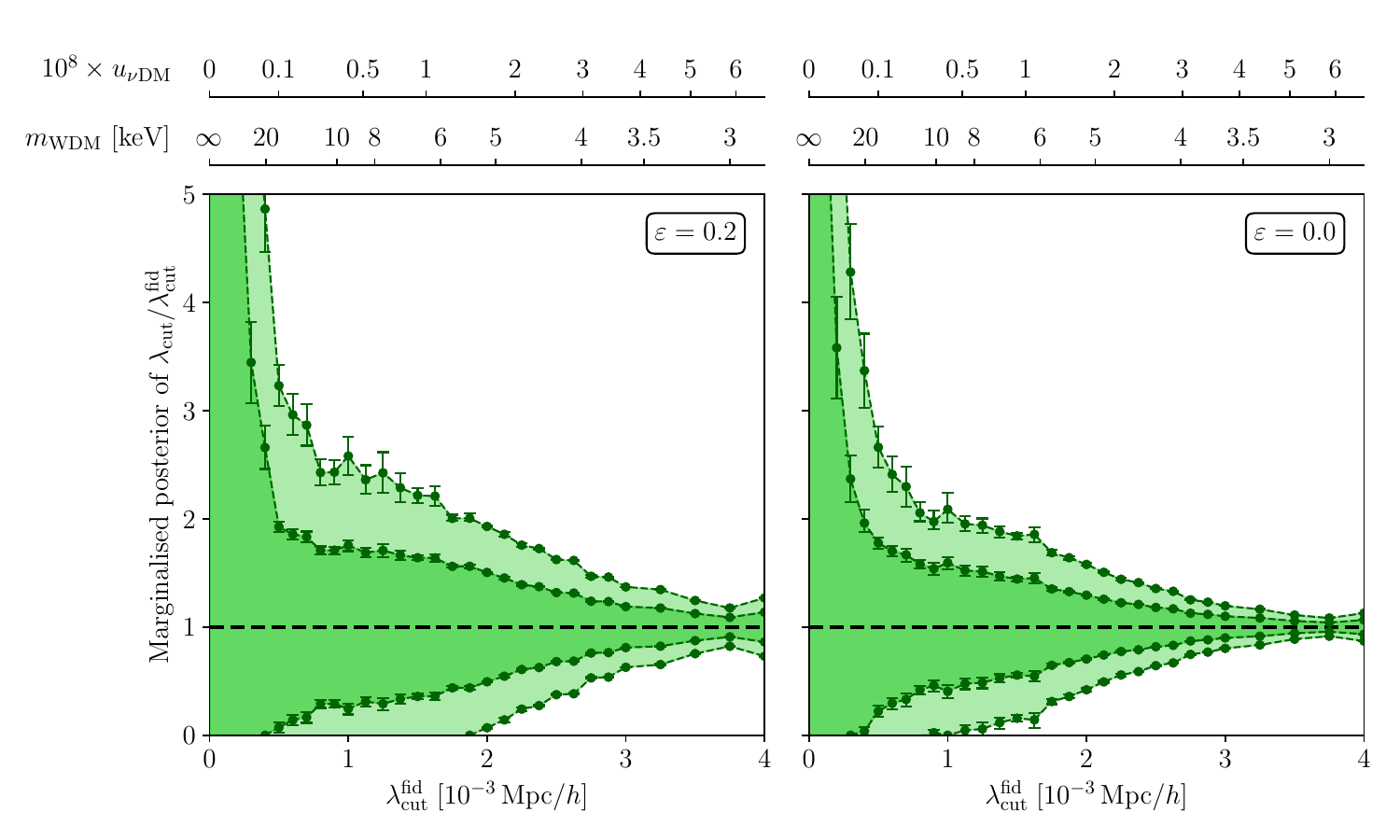}
    \caption{Marginalised $68\%$ (pale shaded area) and $95\%$ (dark shaded area) confidence intervals of $\lcut$, normalised to the value of $\lcutfid$, as a function of $\lcutfid$. The results are obtained by applying the fitted transfer function of Eq.~(\ref{eq:transf_function_fit_form}) to the CDM matter power spectrum (input method~\ref{it:fit}). A conversion between $\lcut$, $m_{\rm WDM}$ and $u_\nudm$ is also provided. The left panel corresponds to $20\%$ modelling noise ($\varepsilon=0.2$) while the right panel assumes no modelling noise ($\varepsilon=0$). The markers show the obtained marginalised posteriors for each value of $\lcutfid$ for which we have effectively performed a Fisher forecast, and the corresponding error bars, see text for details.}
    \label{fig:detection_threshold_plot}
\end{figure}

Figure~\ref{fig:detection_threshold_plot} shows the marginalised posteriors of the reconstructed value of $\lcut$, normalised to the respective fiducial value $\lcutfid$, as a function of $\lcutfid$. The dark shaded areas correspond to 68\%~CL contours while the pale shaded ones denote the 95\% CL contours. To obtain the results, the transfer function fit of Eq.~(\ref{eq:transf_function_fit_form}) is applied directly to the ${\rm CDM}$ matter power spectrum following method \ref {it:fit}. The markers show the obtained constraints at $68\%$ and $95\%$ CL for each evaluated $\lcutfid$ and the error we associate to each confidence interval, following the prescription given in appendix~\ref{app:fisher_plateau}. In the left panel, we assume a 20\% modelling error ($\varepsilon=0.2$), while in the right panel we neglect the modelling error, which greatly improves the value of the detection threshold.

Indeed, with $\varepsilon=0.2$, we find that HERA could make a detection at $95\%$ CL of cutoff scales up to $\lcut\sim 2\times 10^{-3}\,$Mpc/$h$. This detection threshold corresponds to the smallest value of $\lcutfid$ for which the reconstructed $95\%$ confidence interval does not include $\lcut=0$. This corresponds to WDM with a mass of $m_{\rm WDM}=5.1$ keV or $\nudm$ interactions with an interaction strength of $u_\nudm =1.7\times 10^{-8}$. When assuming no modelling error ($\varepsilon=0$), we find that the detection threshold reduces to $\lcut\sim 1\times 10^{-3}\,$Mpc/$h$, which corresponds to $m_{\rm WDM}=9.3\,\rm keV$ or $u_\nudm=4.1\times10^{-9}$. The strong improvement of the detection threshold with decreased modelling error is to be expected. In the redshift bins with $5\lesssim z\lesssim7.5$, the total noise in the lowest, most sensitive $k$-bins is strongly dominated by the modelling noise, which makes any results derived under the assumption of significant modelling noise rather sensitive to the value of $\varepsilon$. We also find that the detection thresholds we obtain are independent of the choice of input method (\ref{it:fit}-\ref{it:nuDM}). This implies that the presence of acoustic oscillations in the $\nudm$ case does not impact the $21\,$cm power spectrum enough to lead to a different detection threshold. We comment further on the ability to distinguish $\nudm$ from WDM in the data in Sec.~\ref{sec:disctinction}.

The 21cm cosmology forecasts are expected to be sensitive to the astrophysical model, see e.g.~\cite{Lopez-Honorez:2016sur, HERA:2021noe, Facchinetti:2023slb, Decant:2021mhj,Greig:2024zso, Agius:2025xbj, Agius:2025nfz, Decant:2024bpg} who studied the impact of the astrophysical modelling onto the 21~signal or onto the reconstruction of the parameters shaping this signal, potentially including exotic DM physics. In particular, increasing $L_X$ and $E_0$ to larger values could give rise to a smoother $21\,$cm signal and thus a more suppressed power spectrum that might become more difficult to probe and make it more difficult to extract DM properties~\cite{ Mesinger:2013nua, Facchinetti:2023slb, Decant:2024bpg}. In this work we find that our results do  not change significantly when varying the X-ray parameters. We have varied $L_X^{\rm II}=L_X^{\rm III}$ between $10^{41}$ and $10^{40} \,{\rm erg\,s}^{-1}\,{\rm M}_\odot^{-1}\,\rm yr$, as allowed by the existing HERA data~\cite{HERA:2021noe, HERA:2022wmy}, and increased $E_0$ up to 1.1 keV.  
While the individual confidence intervals of $\lcut$ for the same $\lcutfid$ varied by up to $\sim15\%$ between models, this did not affect the obtained detection thresholds at the level of the resolution that we have, even in the case where we set $\varepsilon=0$.

In the case of WDM and for $\varepsilon=0.2$, the obtained WDM mass threshold for detection is already disfavoured by Lyman-$\alpha$ data, see~\cite{Irsic:2023equ}  and the discussion in Sec.~\ref{sec:intro}. On the other hand,  in the case of  $\nudm$, the authors of \cite{Hooper:2021rjc} claimed a preference for $\nudm$ interactions with $u_{\rm \nu DM} \sim 5.5 \times 10^{-5}$, corresponding to $\sigma_\nudm\sim 3.7 \times 10^{-32}$ cm$^2$ for DM with a mass $m_{\rm DM}=1\,\rm GeV$, using CMB, BAO and Lyman-$\alpha$ data. This would result in a cutoff scale similar to that of WDM with a mass of $m_{\rm WDM}\sim 0.4$ keV. As this is already strongly excluded by a variety of data,  the authors argued that the extra features in the matter power spectrum imprinted by $\nudm$ interactions could be at the origin for this preference. In our analysis, we find a direct correspondence between the forecasted HERA detection thresholds for $u_{\rm \nu DM}$ and $m_{\rm WDM}$. Therefore, even if HERA should not be able to constrain WDM more than it already is by Lyman-$\alpha$ data when considering 20\% modeling error, it should still improve on the detectability of $\nudm$ with a detection threshold at $u_{{\rm \nu DM}}= 1.7\times 10^{-8}$. As this is orders of magnitude below the asserted preferred value of $u_\nudm$, HERA should thus easily be able to confirm or disprove the claim. 
Accounting for the difference in methodology and telescope, our results are in good agreement with Ref.~\cite{Dey:2022ini}, which reported forecasted constraints at the level of $u_{\rm \nu DM} \le  6.6\times 10^{-7}$ and Ref.~\cite{Mosbech:2022uud}, which estimated constraints of $m_{\rm WDM} \ge 4$ keV and $u_{\rm \nu DM} \le 3.6 \times 10^{-8}$ for 1000 hours of observation time with SKA.  

In this work, in the most optimistic scenario, neglecting any modelling error on the signal ($\varepsilon=0$), we find that HERA forecast should  probe WDM masses up to $ m_{\rm WDM}\sim 9.3$ keV, significantly above the current Lyman-$\alpha$ bound. Bearing in mind the differences in methodology, astrophysical model and potentially $21\,$cm simulator, this is in line with the results of~\cite{Decant:2024bpg,Giri:2022nxq}.  Within the same framework, $\nudm$ interactions could be probed for interaction strengths down to $u_\nudm\sim 4.1 \times 10^{-9}$.

\subsection{Astrophysics-DM Degeneracies }
\label{sec:degen-results}

\begin{figure}
    \centering
    \includegraphics[width=0.65\linewidth]{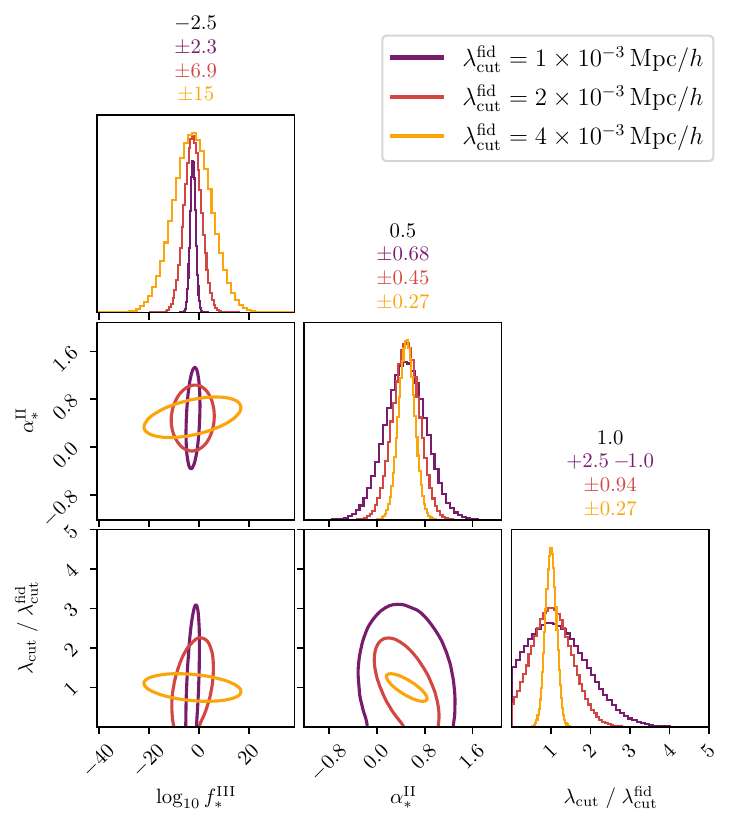}
    \caption{Marginalised one- and two-dimensional posteriors at $95\%$ CL from our Fisher matrix analysis of a reduced sets of parameters for different values of $\lcutfid$, assuming the transfer function as the input method. The one-dimensional posterior of each parameter is shown along the diagonal axis of the plot, and the fiducial value and $95\%$ confidence interval for the respective parameter are written above. The yellow contours correspond to larger values of $\lcutfid$, or equivalently warmer or more strongly interacting DM, while darker contours correspond to colder or more weakly interacting DM.}
    \label{fig:wdm_triangle_plot_small}
\end{figure}

Figure~\ref{fig:wdm_triangle_plot_small} shows the marginalised one- and two-dimensional posteriors at $95\%$ CL for a selection  of astrophysical parameters that are expected to show some level of degeneracy with $\lcut$, see the discussion in Sec.~\ref{sec:effect}. Here again we assume the transfer function as the input method (method~\ref{it:fit} above). For completeness, a full plot containing all parameters is available in the appendix with Fig.~\ref{fig:wdm_full_triangle_plot}. The warmer (colder) or equivalently more strongly (weakly) interacting DM case is shown as a yellow (purple) contour, corresponding to $\lcutfid=4\times10^{-3}\,$Mpc/$h$ ($1\times10^{-3}\,$Mpc/$h$). The intermediate case shown in red corresponds to $\lambda_{\rm cut}=2\times10^{-3}\,$Mpc/$h$, which is the smallest detectable cutoff scale at $\varepsilon=0.2$, as discussed above. 

As can be seen, for larger cutoff scales, or equivalently warmer or more strongly interacting DM (yellow), the constraints on the astrophysical parameters related to MCGs in minihalos become much less stringent. This is well-visible for $f_\star^{\rm III}$ in Fig.~\ref{fig:wdm_triangle_plot_small}, where the posterior widens significantly for larger $\lcut$ (yellow lines). The same behaviour can be observed for $\alpha_\star^{\rm III},\;f_{\rm esc}^{\rm III}$ and $L_X^{\rm III}$ in Fig.~\ref{fig:wdm_full_triangle_plot} in the appendix. This is a result of the suppression induced by the NCDM of low-mass halos and subsequently their impact on the $21\,$cm signal. The uncertainties on ACG parameters in contrast decrease if the NCDM suppression is very strong (large $\lcut$). This is well-visible for $\alpha_\star^{\rm II}$ in Fig.~\ref{fig:wdm_triangle_plot_small}, but also for the ACG parameters $f_\star^{\rm II}$ $f_{\rm esc}^{\rm II}, L_X^{\rm II}$ (as well as for $t_\star$ and $\alpha_{\rm esc}$, which in our astrophysical model are shared between ACGs and MCGs). Conversely, in the "colder" dark matter case (purple), the uncertainty on the MCG parameters is reduced while the reconstruction of $\lambda_{\rm cut}$ worsens. This also shows in the widening of the posterior of $\lcut/\lcutfid$ in Fig.~\ref{fig:detection_threshold_plot} at low $\lcutfid$. There are thus two competing effects: when MCGs contribute the most (small $\lcut$, CDM-like), parameters related to those galaxies can be reconstructed very well and show little degeneracy with $\lcut$, meaning that $\lcut$ can also be reconstructed well. When MCGs barely contribute (at large $\lcut$, NCDM suppressed),  ACG parameters can be reconstructed  better. In both of these cases, degeneracies between some astrophysical parameters and $\lcut$ can be reduced (see the yellow and purple contours in the bottom row of Fig.~\ref{fig:wdm_triangle_plot_small}). 

In the intermediate regime, astrophysical parameters show stronger degeneracies with $\lcut$. This is illustrated with the example of the red curves in Fig.~\ref{fig:wdm_triangle_plot_small}, which correspond to $\lcut=2\times10^{-3}\,{\rm Mpc}/h$. The degeneracies can be understood with the discussion in Sec.~\ref{sec:effect}.  We see for example that $\lambda_{\rm cut}$ is positively correlated with  $f_\star^{\rm III}$, as decreasing the latter parameter reduces the contribution of MCGs to all forms of radiation and pushes all features of the $21\,$cm power spectrum towards later times. This can partially be compensated by a decrease of $\lambda_{\rm cut}$, which leads to colder DM and less suppressed MCGs, see Fig.~\ref{fig:astro-impact-PS-1}. In contrast, $\lambda_{\rm cut}$ is anti-correlated with $\alpha_\star^{\rm II}$. An increase of $\alpha_\star^{\rm II}$ indeed leads to an effective suppression of the contribution of low mass halos w.r.t. to larger ones, see Fig.~\ref{fig:astro-impact-PS-2}, which can then partially be compensated by a decrease in $\lambda_{\rm cut}$.

\subsection{Distinguishing WDM from $\nudm$}
\label{sec:disctinction}

Another relevant question is, assuming that HERA detects a signal compatible with an NCDM model, will it be able to discriminate whether it is WDM or $\nu$DM? Fisher matrices only allow us to compare the different values obtained in the two different configurations (WDM or  $\nudm$) for the detection thresholds of $\lambda_{\rm cut}$. For a given astrophysical model, if these two values are \emph{similar}, one can expect the models to be undistinguishable, none imprinting a distinctive signature. On the contrary, if not \emph{similar}, one would be able to identify the detected model in the window of $\lambda_{\rm cut}$ values were one is detectable while the other is not.  Yet, this simple discussion cannot constitute a rigorous argument. Below we provide a more quantitative assessment.

Let us assume that one NCDM model is the underlying "true" model of the universe, and that a given value $\lcut\neq0$ is detected.  We consider the possibility to  interpret the data within a "test" NCDM model that differs from the "true" model. We want to assess how well the "test" model fits the data and whether it would induce any bias in the reconstruction of $\lcut$. To do so, we compute the quantity
\begin{equation}
   \Delta \chi^2_{\rm test|true}(\lambda_{\rm cut}^{\rm test}, \lambda_{\rm cut}^{\rm true})  \equiv \sum_{i_k, i_z} \left(\frac{{\cal O}^{\rm true}(i_k, i_z \, | \, \lambda_{\rm cut}^{\rm true}) - {\cal O}^{\rm test}(i_k,  i_z \, | \,  \lambda_{\rm cut}^{\rm test})}{\sigma_{21}^{\rm true}(i_k, i_z \, | \,  \lambda_{\rm cut}^{\rm true})} \right)^2 \, ,
\label{chi2_definition}
\end{equation}
where the "true" and "test" models can be either one of WDM and $\nudm$ (denoted as "W" and "$\nu$" for short, respectively), the observable is ${\cal O}=\overline{\delta T}^2_{21} \Delta_{21}^2$, and the measurement error $\sigma_{21}^{\rm true}(i_k, i_z \, | \,  \lambda_{\rm cut})$ is computed in each $k$- and $z$-bin for the true $21\,$cm power spectrum. All model parameters other than 
$\lambda_{\rm cut}$ are set to the values reported in Tab.~\ref{tab:fiducial_astro_parameters}. Although this approach suffers from the same shortcoming as the Fisher matrix formalism in that it assumes a fixed astrophysical fiducial model, it does provide a more concrete estimate of HERA's potential to discriminate between the models.

\begin{figure}[t]
    \centering
    \includegraphics[width=0.6\linewidth]{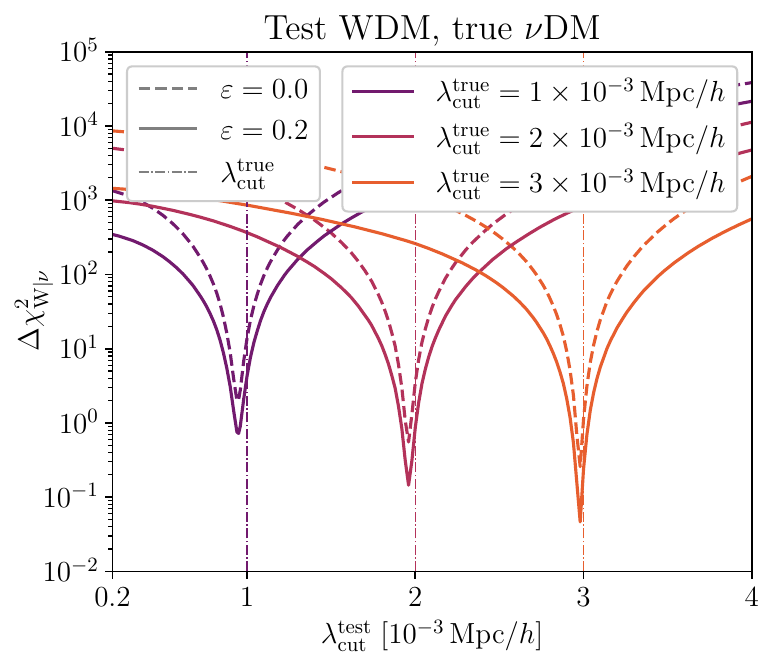}
   
    \caption{
    $\Delta \chi^2_{\rm W|\nu}$, testing the goodness of fit of WDM for an underlying, "true" $\nudm$ model, as a function of the "test" value of $\lcut$ for 3 values of the "true" value $\lcut^{\rm true}$. The case where $\varepsilon=0.2$ is displayed as continuous lines, while  $\varepsilon=0$ is shown with dashed lines.}
    \label{fig:AIC1D}
\end{figure}

In Fig.~\ref{fig:AIC1D} we show the resulting values of $\Delta\chi^2_{W|\nu}$ for three different values of $\lcut^{\rm true}=\{1, 2$ and $3\}\times 10^{-3}\,$Mpc/$h$ (purple, red and orange colours, respectively), where we take $\nudm$ to be the true model and WDM to be the test model. The values of $\lcut^{\rm true}$ are indicated with coloured vertical dot dashed lines. We show  the cases where $\varepsilon=0.2$ (continuous lines) and $\varepsilon=0$ (dashed lines). Both cases show a very similar trend. A larger value of $\varepsilon$ corresponds to more noise and thus less discriminatory ability. Furthermore, the value of $\Delta\chi^2$ decreases with increasing cutoff scale $\lcut^{\rm true}$, which means that the discriminatory ability between WDM and $\nudm$ decreases when DM is warmer or more strongly interacting. For completeness, we show a full 2D plot of $\log_{10}\Delta\chi^2_{{\rm W}|\nu}$ as a function of $\lcut^{\rm true}$ and $\lcut^{\rm test}$ in the appendix in  Fig.~\ref{fig:AIC_2D}.

One way to quantify the discriminatory ability of HERA is to compute the difference in the Akaike Information Criterion (AIC) between the true and the test model. The AIC is defined as ${\rm AIC}\equiv2 N_{\rm DOF}-2\ln\mathcal{\hat L}$, where $\mathcal{\hat L}$ is the maximum likelihood and $N_{\rm DOF}$ the number of degrees of freedom of the model. The relative likelihood of the test model compared to the true model is then given by $\exp(-\Delta{\rm AIC_{\rm test|true}}/2)$. Under the assumption of a Gaussian likelihood and given that in our case, the true model perfectly fits the data, the AIC difference reduces to $\Delta{\rm AIC}_{\rm test|true}=\min\Delta \chi^2_{\rm test|true}$. We find that already when keeping the astrophysical parameters fixed, $\min\Delta\chi^2_{\rm test|true}\lesssim\mathcal{O}(1)$ in all cases. This corresponds to relative likelihoods of the test models of $\gtrsim0.6$, which indicates no statistically meaningful discrimination. Allowing the astrophysical parameters to vary would only further weaken HERA’s ability to distinguish between these models.

In both Figs.~\ref{fig:AIC1D} and~\ref{fig:AIC_2D}, it is visible that for values $\lcut^{\rm true}\lesssim1\times10^{-3}\,{\rm Mpc}/h$, the best-fit of $\lcut^{\rm test}$ is at a value of $\lcut<\lcut^{\rm true}$. This means that for such values of $\lcut^{\rm true}$, assuming WDM in place of $\nudm$ would yield a biased reconstruction of $\lcut$ towards smaller values. This can be understood in that the $\nudm$ model gives rise to more power (and thus more formation of structures) at small scales than WDM, making it more CDM-like for the same value of $\lcut$ and inducing a smaller shift in the $21\,$cm power spectrum towards later redshifts than the equivalent WDM model does. The fact that the bias disappears for larger $\lcut^{\rm true}$ could be understood in that for such large values of $\lcut^{\rm true}$, the MCG population is so suppressed that it cannot significantly impact the $21\,$cm signal, and no difference appears. Yet, in all cases represented here the bias is small.

\section{Conclusion}
\label{sec:conclusion}

In this paper, we have explored the sensitivity of the near-future radio telescope HERA, that probes the fluctuations in the $21\,$cm signal, to two different non-cold dark matter (NCDM) scenarios. Here we focus on thermal warm dark matter (WDM) and dark matter interacting with neutrinos ($\nudm$), two NCDM models which exhibit similar but not equal suppressions of small-scale structure formation. The suppressions are sourced by different physical mechanisms: free-streaming in the case of WDM and collisional damping with  acoustic oscillations for $\nudm$. 
In Sec.~\ref{sec:ncdm}, we discuss the WDM and $\nudm$ power spectra and the associated transfer functions, which can be fit in terms of a cutoff scale $\lcut$ that directly depends on the NCDM parameters. These are namely the WDM mass, $m_{\rm WDM}$, and the neutrino-DM interaction strength, $u_\nudm$. We took special care to self-consistently account for massive neutrinos. The definition of the cutoff scale allows us to map $\nudm$ and WDM onto each other, a process that neglects the extra power on small scales caused by the dark acoustic oscillations in the $\nudm$ case.

In Secs.~\ref{sec:HMF} and~\ref{sec:21cm}, we detail how we derive the halo mass function and the power spectrum of $21\,$cm signal fluctuations from the linear matter power spectrum considering two populations of galaxies, namely ACGs and MCGs. Our reference astrophysical scenario is based on the EOS2021 best-fit parameters of Ref.~\cite{Munoz:2021psm} and is given in Tab.~\ref{tab:fiducial_astro_parameters}.  We make use of the extended Press-Schechter formalism, using a sharp-$k$ window function, which is well-suited for NCDM scenarios.  In the subsequent description of our astrophysical model, we have  introduced the stellar mass distribution, which depends on both the halo mass function and the efficiency of star formation. We have shown that in the presence of NCDM, the MCG contribution is strongly affected by the NCDM-induced damping of small-scale structures. This is because MCGs are hosted in minihalos, the abundance of which is significantly impacted by the NCDM. We further detail the impact of NCDM on the $21\,$cm signal. As already noted in previous works, NCDM tends to shift the signal to lower redshift, as all forms of radiation fluxes  shaping the signal are suppressed. The $\nudm$ scenario induces a slightly less pronounced shift than WDM due to the extra power on small scales, yet this effect appears to be minor for the considered astrophysical framework. We also discuss some of the degeneracies with astrophysical parameters.

In Sec.~\ref{sec:analysis}, we employ the Fisher matrix formalism to forecast HERA's sensitivity to the NCDM scenarios, assuming 1000 hours of observation, and determine the detection threshold for $\lcut$. We find that the results are independent of our input method, either using the WDM ($\nudm$) power spectrum from {\tt CLASS} with the corresponding $m_{\rm WDM}$ ($u_\nudm$) as the input parameter, or applying directly the featureless fit of the transfer function to the CDM power spectrum. We also show  that the detection threshold strongly depends on the amount of modelling noise that we assume. For the standard $20\%$ modelling noise, we find that HERA could detect values of $\lcut\approx 2\times10^{-3}\,{\rm Mpc}/h$ at $95\%$ CL, which corresponds to $m_{\rm WDM}\approx5.1\,\rm keV$ or $u_\nudm\approx1.7\times10^{-8}$. These thresholds are either already excluded by  current Lyman-$\alpha$ constraints in the WDM case or  will allow HERA to test recent claims of a non-zero interaction strength \cite{Hooper:2021rjc} in the $\nudm$ case. Under the assumption of no modelling noise, the detection threshold improves to $\lcut\approx 1\times10^{-3}\,{\rm Mpc}/h$ at $95\%$ CL, which corresponds to $m_{\rm WDM}\approx9.3\,\rm keV$ or $u_\nudm\approx4.1\times10^{-9}$. In this optimistic case, HERA would be able to probe the NCDM scenarios well beyond e.g. current Lyman-$\alpha$ limits. We find that these results barely change when varying the heating parameters.

We also discuss the degeneracies between NCDM and astrophysical parameters in the Fisher matrix posteriors, recovering the expected dependencies as discussed in Sec.~\ref{sec:21cm}. We find that when MCGs are strongly suppressed, ACG parameters are well-reconstructed and degeneracies between those and $\lcut$ are alleviated. When MCGs are less suppressed (CDM-like), MCG parameters are well-reconstructed with little degeneracy with $\lcut$, at the expense of ACG parameters. In the transitional regime, neither ACG nor MCG parameters can be reconstructed as precisely and we observe stronger degeneracies between $\lcut$ and astrophysical parameters. Finally, we more quantitatively study the capacity of HERA to distinguish between $\nudm$ and WDM through a $\Delta \chi^2$ test. We confirm that the expected bias in $\lcut$ is non-zero but small when assuming the wrong model in the reconstruction of the cutoff scales for the value range of interest.  

Our results show that, assuming two populations of galaxies, the $21\,$cm signal has the power to further probe the properties of NCDM. The constraining power will significantly depend on the amount of modelling noise that can be reached in the $21\,$cm signal simulations. Yet, even in the most optimistic scenarios with negligible modelling uncertainty, the extra power on small scales in the $\nudm$ case will unfortunately not allow this model to be distinguished from WDM.

\appendix
\section{Cosmology, NCDM and Astrophysics details}
\label{app:cosmo}

Throughout this paper, we use the natural units convention ($\hbar = k_{\rm B} = c=1$) and a default cosmology from Planck 2018 including TT,TE,EE+lowE+lensing+BAO \cite{Planck:2018vyg}, ($h=0.6766$, $\Omega_{\rm b}h^2 = 0.02242$, $\Omega_{\rm c}h^2 = 0.11933$, $\sigma_8 = 0.8102$, $n_{\rm s} = 0.9665$). In addition, we work with the most up-to-date value of the effective number of light degrees of freedom $N_{\rm eff} = 3.044$ \cite{Froustey:2020mcq, Bennett:2020zkv}. Fisher analyses are produced with {\tt 21cmCAST}\footnote{\url{https://github.com/gaetanfacchinetti/exo21cmCAST}}. We use the following parameters for {\tt 21cmFAST}:
\texttt{redshift}~$=3.5$, 
\texttt{max\_redshift}~$=35$, 
\texttt{HII\_DIM}~$=200$, 
\texttt{BOX\_LEN}~$=300$, 
\texttt{DIM}~$=800$, 
\texttt{PERTURB\_ON\_HIGH\_RES}~$=\texttt{True}$, 
\texttt{PS\_FILTER}~$=\texttt{SHARPK}$, 
\texttt{USE\_RELATIVE\_VELOCITIES}~$=\texttt{False}$,\footnote{One can account for star formation quenching in the smallest galaxies (in MCGs) due to the DM-baryon relative velocities, see~\cite{Munoz:2019hjh}. Yet,  this would require a consistent implementation of relative velocity feedback in a NCDM scenarios. We do not address this technical issue in our work and select the option   {\tt USE\_RELATIVE\_VELOCITIES $=$ False} in {\tt 21cmFAST}.} \texttt{USE\_MINI\_HALOS}~$=\texttt{True}$, 
\texttt{USE\_MASS\_DEPENDENT\_ZETA}~$=\texttt{True}$, 
\texttt{SUBCELL\_RSD}~$=\texttt{True}$, 
\texttt{INHOMO\_RECO}~$=\texttt{True}$, 
\texttt{USE\_TS\_FLUCT}~$=\texttt{True}$. We also perform all CLASS calculations with the in-built fluid approximation disabled, i.e. by setting {\tt fluid\_approximation} $= 3$. Furthermore, note that in this paper  the CDM model refers to $\Lambda{\rm CDM}$ with 3 massive neutrinos and that in all cosmologies considered here the three neutrino masses have been set to $m_\nu=0.02$ eV.

\subsection{Free-streaming length}
\label{app:free_stream}

Here we detail  how the approximation of Eq.~(\ref{eq:k_free_streaming}) for the free-streaming length of the WDM has  been derived. In particular, its definition in Eq.~(\ref{eq:lamfs}) is directly related to the induced cutoff scale in the linear matter power spectrum.  After decoupling, the momentum of one WDM particles follows the FLRW geodesics and evolves as  $p(a) = p_{\rm kd} a_{\rm kd}/a$. Replacing this expression into the definition of the comoving free-streaming length yields
\begin{equation}
    \lambda_{\rm fs} =\int_{a_{\rm kd}}^a \frac{{\rm d}a'}{a'^2H(a')}\frac{\frac{p_{\rm kd}}{T_{\rm kd}}}{\sqrt{a'^2 \left(\frac{m}{a_{\rm kd}T_{\rm kd}}\right)^2 + \left(\frac{p_{\rm kd}}{T_{\rm kd}}\right)^2}} \, ,
\end{equation}
where we have introduced the temperature at kinetic decoupling $T_{\rm kd}$. This temperature satisfies, by definition, $T_{\rm d, 0} \equiv T_{\rm kd} a_{\rm kd}$. From the new expression of $\lambda_{\rm fs}$ and for simplicity, in the following we will use the reduced mass and momenta
\begin{equation}
    \mu \equiv \frac{m}{T_{\rm d, 0}}\, , \quad {\rm and} \quad q \equiv \frac{p_{\rm kd}}{T_{\rm kd}}\, .
\end{equation}
To follow the computation further, one first needs the initial condition at kinetic decoupling of the WDM particle dynamics. The distribution of momenta at kinetic decoupling is given by the Fermi-Dirac distribution, which in terms of the variable $q$ can be written
\begin{equation}
    p_q(q) = \frac{1}{N}\frac{q^2}{e^{\sqrt{(a_{\rm kd}\mu)^2 + q^2}}+1} \, ,
\end{equation}
where $N$ is a normalisation constant. By nature, we must assume that almost all the WDM particles are relativistic when the entire fluid kinetically decouples. This implies that $T_{\rm kd} \gg m$, which, put differently, is $a_{\rm kd} \mu \ll 1$. Thus, this leads, in excellent approximation, to 
\begin{equation}
    p_q(q) \simeq \frac{2}{3\zeta(3)} \frac{q^2}{e^q +1} \, .
\end{equation}
Then, to evaluate $\lambda_{\rm fs}$ at a given WDM mass, one can consider the typical free streaming length defined from the the average momentum as
\begin{equation}
       \overline \lambda_{\rm fs} \simeq \int_{0}^a \frac{{\rm d}a'}{a'^2H(a')}\frac{\left<q\right>}{\sqrt{(a'\mu)^2 + \left<q\right>^2}} \, , \quad {\rm with} \quad \left< q \right> = \int_0^\infty p_q(q) q {\rm d}q = \frac{7\pi^4}{180\zeta(3)} \, .
\end{equation}
In addition, plugging Eq.~(\ref{eq:Td0}) into the expression of $\mu$ one obtains
\begin{equation}
    \mu = 2.3 \times 10^{8}\left(\frac{0.12}{\Omega_{\rm DM}h^2}\right)^{1/3} \left(\frac{2.73 ~{\rm K}}{T_{\rm \gamma, 0}}\right)\left(\frac{m_{\rm WDM}}{5~{\rm keV}}\right)^{4/3} \, .
\end{equation}
Therefore, for $m_{\rm WDM} > 5$ keV one has $\mu > 10^8$, while $\left< q \right> \simeq 3.15$. Consequently, the integral in the expression of $\lambda_{\rm fs}$ is largely dominated by the first part, in range of scale factors $a'$ corresponding to the radiation dominated era (i.e. for small values of $a'$ in the square root in the denominator). A good approximation for $\lambda_{\rm fs}$ is thus 
\begin{equation}
\begin{split}
    \overline \lambda_{\rm fs} & \simeq \frac{\left<q\right>}{H_0 \sqrt{\Omega_{\gamma, 0}}} \int_0^{a_{\rm eq}}  {\rm d} a' \frac{1}{\sqrt{(a'\mu)^2 + \left<q\right>^2}}  \\
    & \simeq   \frac{\left<q\right>}{H_0 \sqrt{\Omega_{\gamma, 0}}} \frac{1}{\mu} \left\{\ln \mu + \ln \left(\frac{2a_{\rm eq}}{\left<q\right>}\right) \right\} \, .
    \end{split}
\end{equation}
The typical comoving free streaming mode is then
\begin{equation}
    \overline{k}_{\rm fs} \simeq \frac{2\pi}{\overline{\lambda}_{\rm fs}} \simeq \frac{2 \pi}{\left< q \right>} H_0 \sqrt{\Omega_{\gamma, 0}} \frac{\frac{m}{T_{\rm d, 0}}}{  \ln \left(\frac{m}{T_{\rm d, 0}}\right) + \ln \left(\frac{2a_{\rm eq}}{\left<q\right>}\right) } \, ,
\end{equation}
and plugging in numbers yields
\begin{equation}
    \overline{k}_{\rm fs} \simeq \frac{4.1  r}{1+0.15 \ln r } ~ h / {\rm Mpc} \quad {\rm with} \quad r = \left(\frac{m}{{\rm keV}}\right)\left(\frac{T_{ \gamma, 0}}{T_{\rm d, 0}}\right) \, .
\end{equation}

\subsection{Global fit of the transfer function}
\label{app:fit_TF}

In order to perform a global fit of the transfer function on both the WDM and $\nudm$ scenarios we minimise the total square error
\begin{equation}
\begin{split}
    \mathcal{E}  \equiv  & \sum_{j, l} \left\{ T_{\rm WDM}(k_j,z=0) - f\left(k_{j}, a_{\rm WDM }\left[\frac{1~{\rm keV}}{m_{{\rm DM}, l}}\right]^{b_{\rm WDM}}\right) \right\}^2 \\
    & + 
    \sum_{j, l} \left\{ T_\nudm(k_{j},z=0) - f\left(k_j, a_\nudm\left[\frac{u_{\nudm, l}}{8.5 \times 10^{-7}}\right]^{b_\nudm}\right) \right\}^2 \, .
    \end{split}
    \label{eq:transfer_function_fit_total_relative_error}
\end{equation}
where $m_{{\rm DM}, l}$ corresponds to a grid of 20 points, chosen such that the inverse values $1/m_{{\rm DM}, l}$ are equally spaced, between 1 keV to 20 keV and $u_{\nu{\rm DM}, l}$ is a grid of 20 points logarithmically distributed from $10^{-9}$ to $10^{-6}$. In order to only capture the first break in the slope (and not the subsequent small \emph{bumps}), the arrays $k_{j}$ are limited to the initial range where $T(k_{j, l}) > 0.15$ for each DM mass and DM-neutrino coupling strength. 

The best-fit is found by minimising Eq.~(\ref{eq:transfer_function_fit_total_relative_error}). In the right panel of Fig.~\ref{fig:MPS&TF-LCDM+3m_nu}, we plot the transfer functions obtained numerically from {\tt CLASS} for WDM (solid coloured lines) and $\nudm$ (dashed lines) together with the fitted expression $f(k,\lambda_{\rm cut})$ from Eq.~(\ref{eq:transf_function_fit_form}) (dotted lines). For the fitted transfer function, we use the derived best-fit values of the parameters shown in Tab.~\ref{tab:fitting_function_best_fit_values}. This clearly illustrates the goodness of the fit.

\subsection{Excursion set formalism and the window function choice: top-hat vs sharp-$k$}
\label{sec:HMFapp}

\subsubsection{Excursion - set theory in a nutshell}
\label{sec:excursion}
Assuming Gaussian distributed initial perturbations, the excursion set theory relates the linear matter power spectrum to the halo mass distribution \cite{Bond:1990iw, Zentner:2006vw, GalaxyFormation}. To each region of size $R$ we can associate a smoothed linear density contrast $\delta_R$ over a window $W_R^w$ where $w$ tags the different possible window choices, some of which are discussed below. More precisely, we define
\begin{equation}
    \delta_R^w  \equiv \int {\rm d}^3 {\bf x} \,  \delta({\bf x}) W_R^w(|{\bf x}|)  =  \int \frac{{\rm d}^3 {\bf k}}{(2\pi)^3} \hat \delta ({\bf k}) \hat W_R^w(k)
    \label{eq:def_delta_R}
\end{equation}
where $\hat \delta$ (resp. $\hat W_R^w$)  is the Fourier transform of the density contrast field $\delta({\bf x})$ (resp. $W_R^w$). The first natural choice of window function is the \emph{top-hat}, $W_{R}^{\rm TH}({\bf x}) = 3\Theta(R-|{\bf x}|)/(4\pi R^3)$, which has the Fourier transform
\begin{equation}
\begin{split}
    \hat W_R^{\rm TH}(k) & =  \frac{3}{4\pi R^3} \int_{|{\bf x}| < R} {\rm d}^3 {\bf x} \, e^{-i{\bf k}{\bf x}} \\
    & = \frac{3}{(kR)^3} \left[\sin (kR) - kR \cos (kR)\right] \, ,
    \end{split}
    \label{eq:fourier_top_hat}
\end{equation}
and that defines $\delta_R^{\rm TH}$ as the volume average of the density contrast inside a sphere of radius $R$. Assuming that the Universe is homogeneous and isotropic, and that the linear density contrast is Gaussian-distributed, it follows the probability density functional
\begin{equation}
    \mathcal{P}_\delta[\delta] \propto \exp\left(-\frac{1}{2}\int \frac{{\rm d}^3 {\bf k}}{(2\pi)^3} \frac{|\hat \delta ({\bf k})|^2}{P(k)}\right) \, ,
    \label{eq:PDFunctional}
\end{equation}
where the variance $P(k)$ is the linear matter power spectrum. The smoothed density contrast $\delta_R$ being defined as an integral over a range of modes weighted by the window function, it follows a Gaussian distribution of mean 0 and  variance $S_R^w$ of Eq.~(\ref{eq:variance}). \footnote{Note that Eq.~(\ref{eq:PDFunctional}) gives a implicit definition of the matter power spectrum as 
\begin{equation}
    \left< \hat\delta({\bf k}) \hat\delta^\star({\bf k'}) \right> \equiv (2\pi)^3 \delta_{\rm D}({\bf k} - {\bf k'})P(k)
\end{equation}}

In the excursion set approach, counting halos inside a given volume and around a given point in space can be done by tracking the evolution of $\delta_R$ (until it reaches the linear collapse threshold $\delta_{\rm c}$) as we decrease $R$ -- or equivalently as we increase $S^w = S_R^w$ since there is a one-to-one correspondence between $R$ and $S_R^w$. Let us consider two regions of size $R$ and $R-\epsilon$ with $0 < \epsilon \ll R$. The difference of density contrast between these two regions,
\begin{equation}
    \delta_{R - \epsilon}^w - \delta_R^w = \int \frac{{\rm d}^3 {\bf k}}{(2\pi)^3} \hat \delta ({\bf k}) \left[\hat W_{R-\epsilon}^i(k) - \hat W_R^w(k)\right] \, ,
    \label{eq:DeltaDiff}
\end{equation}
is in general not independent on $\delta_{R}^w$ because the support of $\hat W_{R-\epsilon}^w(k) - \hat W_R^w(k)$ overlaps that of $\hat W_R^w(k)$. This can be seen, in the case of the top-hat, directly from equation~(\ref{eq:fourier_top_hat}). This makes the construction of an analytical theory of structure formation with an arbitrary window function extremely challenging. The only exception is for the sharp-$k$ window function defined from its Fourier transform in Eq.~(\ref{eq:SK}).
With that choice, Eq.~(\ref{eq:DeltaDiff}) reduces to an integral of $\hat \delta({\bf k})$ over $k \in [1/R, 1/(R-\epsilon)]$, completely independent of the associated $\delta_R^{\rm SK}$ obtained from the integral in Eq.~(\ref{eq:def_delta_R}) performed over $k \in [0, 1/R]$. The evolution of $\delta_R^{\rm SK}$ with $R$ can then described as a random walk, each step size being drawn out of a Gaussian distribution $\delta_{R - \epsilon}^{\rm SK} - \delta_R^{\rm SK} \sim \mathcal{N}(0, S_{R-\epsilon}^{\rm SK} - S_R^{\rm SK})$. In the limit $\epsilon \to 0$, $\delta_R^{\rm SK}$ thus follows a Brownian motion and properties of collapsed halos can be directly inferred from its properties \cite{Bond:1990iw}.

\begin{figure}[ht]
    \centering
    \includegraphics[width=0.49\linewidth]{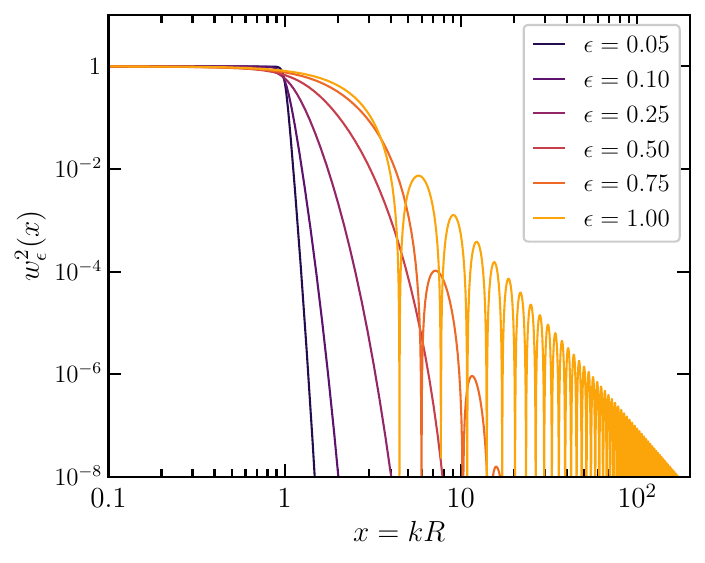}
    \includegraphics[width=0.49\linewidth]{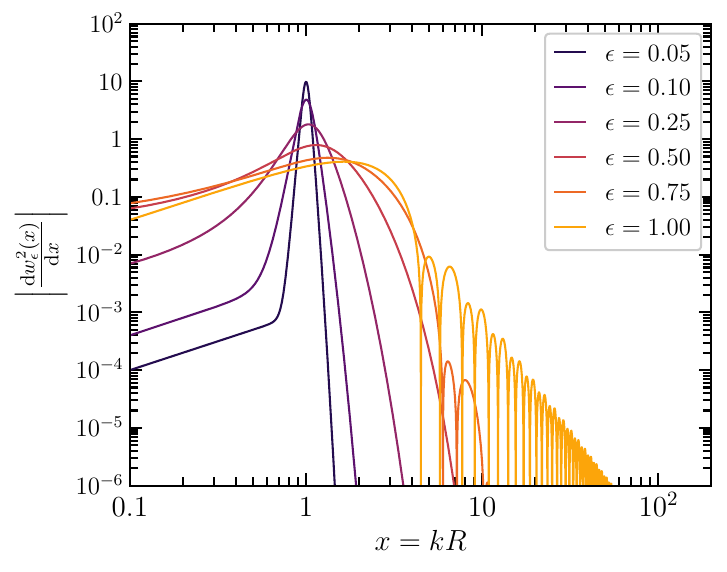}
    \caption{{\bf Left panel}. Square of the generalised window function $w_\epsilon(x)$ for selected values of $\epsilon$. {\bf Right panel}. Absolute value of its derivative, which enters the definition of $S’(R)$.}
    \label{fig:window_function}
\end{figure}

\subsubsection{The importance of the window function}
\label{app:derivative_variance}

In order to recover the correct mass function it is necessary to properly evaluate the derivative of the variance of the matter power spectrum $S$,
\begin{equation}
    S'(R) \equiv \frac{{\rm d}S}{{\rm d} R} = \int \frac{{\rm d}^3 {\bf k}}{(2\pi)^3} P(k) \frac{{\rm d} |\hat W_R(k)|^2}{{\rm d}R} \, .
\end{equation}
In the following we introduce the dimensionless matter power spectrum $\mathcal{P}_{\rm m}(k) = k^3 P(k) / (2\pi^2)$. Moreover, because the Fourier transform of the smoothing window is a function of the product $kR$, it is possible to introduce the reduced window function $w : x \mapsto w(x) = \hat W_R(x/R)$, independent of $R$, such that the variance takes the simple form
\begin{equation}
    S'(R)  = \frac{1}{R} \int_0^{\infty} {\rm d} x \, \mathcal{P}_{\rm m}\left(\frac{x}{R}\right) \frac{{\rm d } w^2(x)}{{\rm d} x} \, .  
\end{equation}
Depending on the shape of the window function this integral may give a different result. To quantify this effect, let us define the family of valid window functions,
\begin{equation}
    w_\epsilon : x \mapsto     w^{\rm TH}(\epsilon x) \left[\frac{1 + \exp \left(-\frac{1}{\epsilon}\right)}{1+\exp\left(\frac{x-1}{\epsilon}\right)}\right]^{1-\epsilon}
\end{equation}
parametrised by $\epsilon \in (0, 1]$, and where $w^{\rm TH}$ is the usual top-hat window function
\begin{equation}
    w^{\rm TH} : y \mapsto \frac{3}{y^3} \left[\sin(y) - y \cos (y)\right] \, .
\end{equation}
This set of functions is chosen to provide a continuous interpolation between the top-hat and the sharp-$k$, $w^{\rm SK} : x \mapsto \Theta(1-x)$, window functions as they satisfy the following limits
\begin{equation}
    \begin{cases}
        w_1(x) = w^{\rm TH}(x)\\[8pt]
        \displaystyle \lim_{\epsilon \to 0} w_\epsilon(x) = w^{\rm SK}(x) 
    \end{cases}
    \quad \forall x \in [0, \infty) \, .
\end{equation}

In the left panel of Fig.~\ref{fig:window_function} we show $w_\epsilon^2(x)$ for selected values of $\epsilon$, while the right panel displays the absolute value of its derivative, which enters the expression of $S’(R)$. As $\epsilon$ decreases from 1 to 0, the oscillatory pattern characteristic of the top-hat window function gradually vanishes, and for small $\epsilon$ the derivative tends toward a Dirac delta distribution. For $\epsilon \sim 1$, the broad derivative implies that the integral for $S’(R)$ receives significant contributions from values of the matter power spectrum at $k \ll 1/R$. When the power spectrum is truncated, this can produce an artificially large $S’(R)$ on scales where the spectrum is nearly zero. Consequently, the value of $S’(R)$ becomes sensitive to the detailed shape of the window function derivative itself, which lacks physical meaning.

\begin{figure}[t]
    \centering 
    \includegraphics[width=0.49\textwidth]{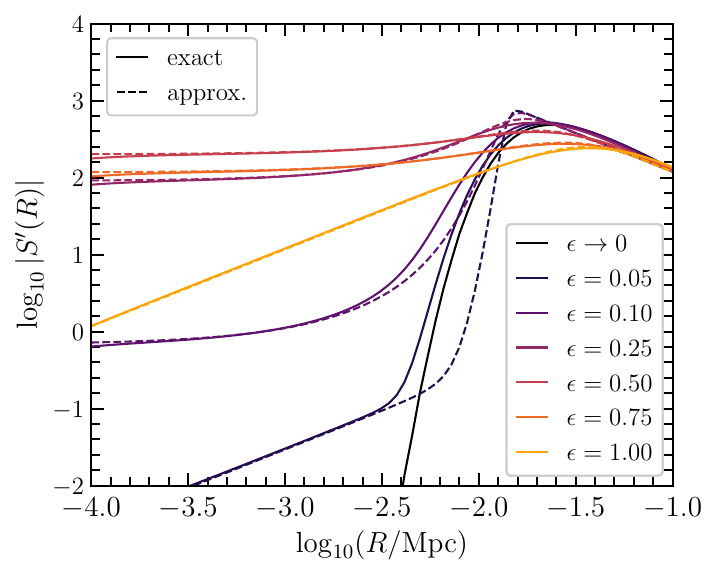}
    \includegraphics[width=0.49\textwidth]{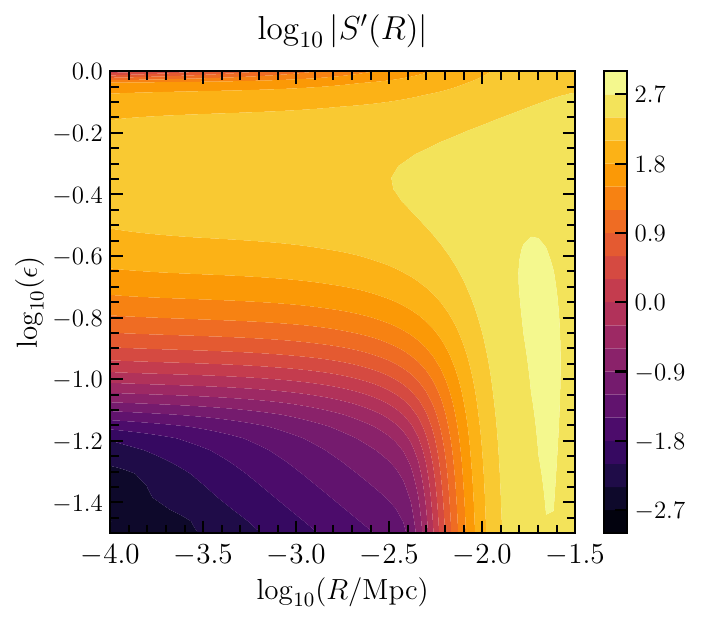}
    \caption{Derivative of the variance of the matter power spectrum in terms of the smoothing radius $R$ for various values of $\epsilon$ (i.e., for different shape of the window function). {\bf Left panel}. Result for a finite set of values for $\epsilon$, comparing the exact expression (solid lines) to the approximation (dashed lines) obtained from Eq.~(\ref{eq:SprimeAnal}). {\bf Right panel}. Contour map in terms of both $R$ and $\epsilon$.}
    \label{fig:error_WindowFunction}
\end{figure}

To illustrate the significance of this issue, Fig.~\ref{fig:error_WindowFunction} shows the behavior of $S’(R)$ as a function of $R$ and $\epsilon$: for selected $\epsilon$ values in the left panel and as a contour map in the right panel. In the left panel, the solid lines correspond to the exact integration using the matter power spectrum of a $\sim 7\,$keV WDM model (computed from the analytical transfer function) displaying suppressed power for wave modes larger than  $k_0 \sim 7\times 10^1\,$Mpc$^{-1}$. The dashed lines, in contrast, assume that the dimensionless matter power spectrum is approximately constant over the relevant modes and sharply truncated at $k=k_0$. The latter case approximation is useful as it allows to derive an analytic expression of the derivative of the variance given by
\begin{equation}
S’(R) \sim \frac{\mathcal{P}_0}{R}\left[ 1- w_{\epsilon}^2(k_0R)\right] \, ,
\label{eq:SprimeAnal}
\end{equation}
with $\mathcal{P}_0$ a normalisation constant (that should be set here to 12 to approximate the $7\,$keV WDM case). Remarkably, this approximation reproduces the exact result particularly well for large $\epsilon$, highlighting the direct correspondence between the shape of $S’(R)$ and that of the window function. The right panel further shows that for $\epsilon \gtrsim 0.1$, $S’(R)$ remains sizeable down to $R < 10^{-4}\,$Mpc, even though the matter power spectrum is sharply truncated around $1/k_0 \sim 10^{-2}\,$Mpc. Moreover, in that case, the precise amplitude depends sensitively on the arbitrary choice of $\epsilon$.

In the end, in the main text, we have first re-derived that the vanilla Press-Schechter halo mass function can only be derived assuming a sharp-$k$ window function. Then, we have shown here that the halo mass function of a truncated matter power spectrum obtained with the top-hat window function is driven by the shape of the window function itself. Therefore, we conclude that the sharp-$k$ window function is the best choice for consistency and accuracy for our models.

\section{Effect of WDM vs $\nu$DM}
\label{sec:WvsN}

\begin{figure}[ht]
    \centering
    \includegraphics[width=0.8\linewidth]{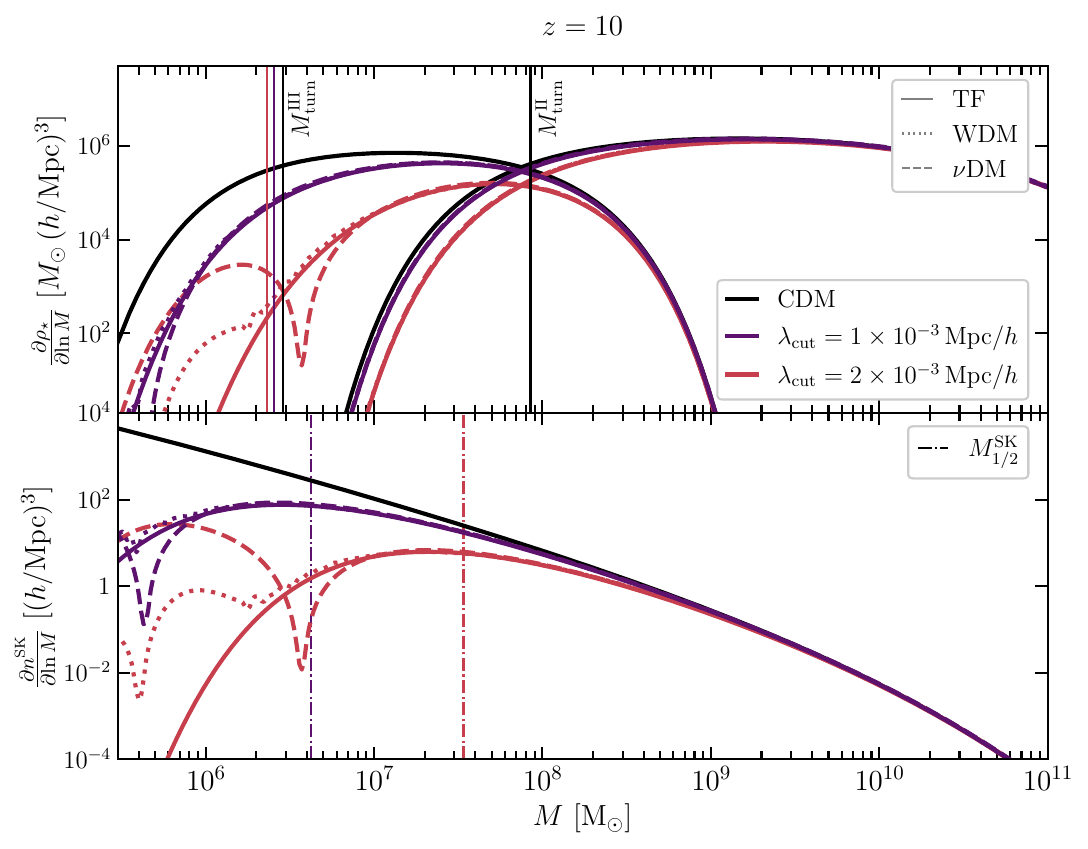}
    \caption{Same as Fig.~\ref{fig:stellar_density_FR}, but for two fixed values of $\lcut$, comparing the effect that the assumption of the fitted transfer function (continuous),  WDM (dotted) or $\nudm$ (dashed) (input methods~\ref{it:fit}-\ref{it:nuDM} respectively) have on the stellar mass distribution (upper panel) and halo mass function (lower panel). For the stellar mass distribution, the halos with masses in the range $10^5\le M/{\rm M}_\odot\le10^9$ get contributions from MCGs while halos in the mass range $10^7\le M/{\rm M}_\odot\le10^{11}$ get contributions from ACGs.}
    \label{fig:hmf_model_comparison}
\end{figure}

Here we provide some extra plots illustrating the differences or similarities between WDM and $\nudm$. The impact of assuming either the WDM or $\nudm$ power spectra or the fitted transfer function as input method~\ref{it:fit}-\ref{it:nuDM} on the halo mass function and stellar mass distribution (discussed in Secs.~\ref{sec:HMF} and~\ref{sec:SFR-ion-Xrays}) is shown in Fig.~\ref{fig:hmf_model_comparison}. We see that the strongest differences start to appear at low halo masses, which are already significantly suppressed w.r.t. the CDM prediction.

\begin{figure}[htb]
    \centering
    \includegraphics[width=0.49\linewidth]{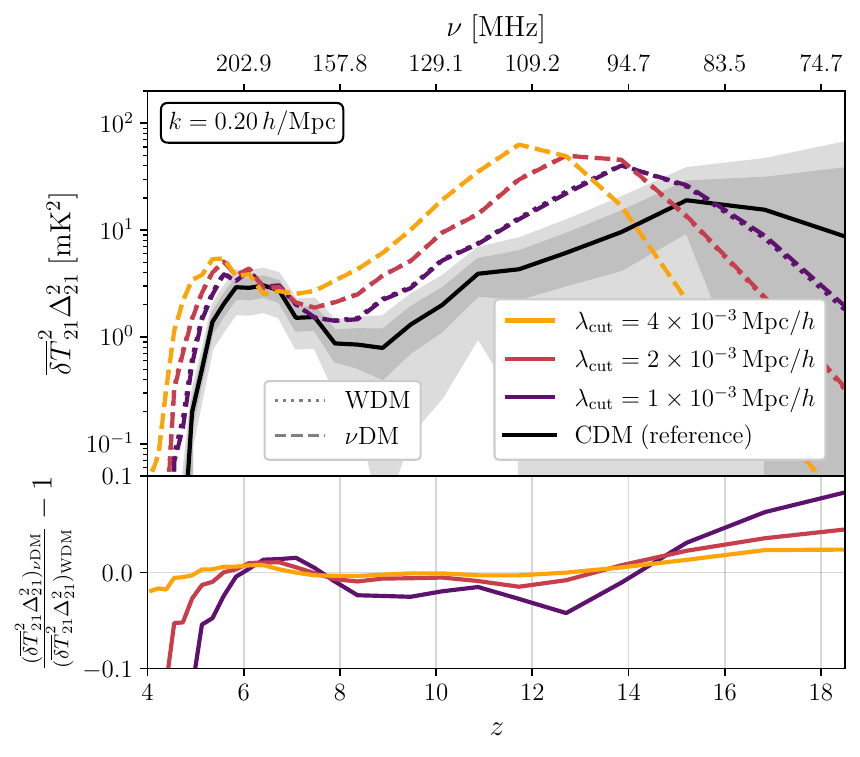}
    \includegraphics[width=0.49\linewidth]{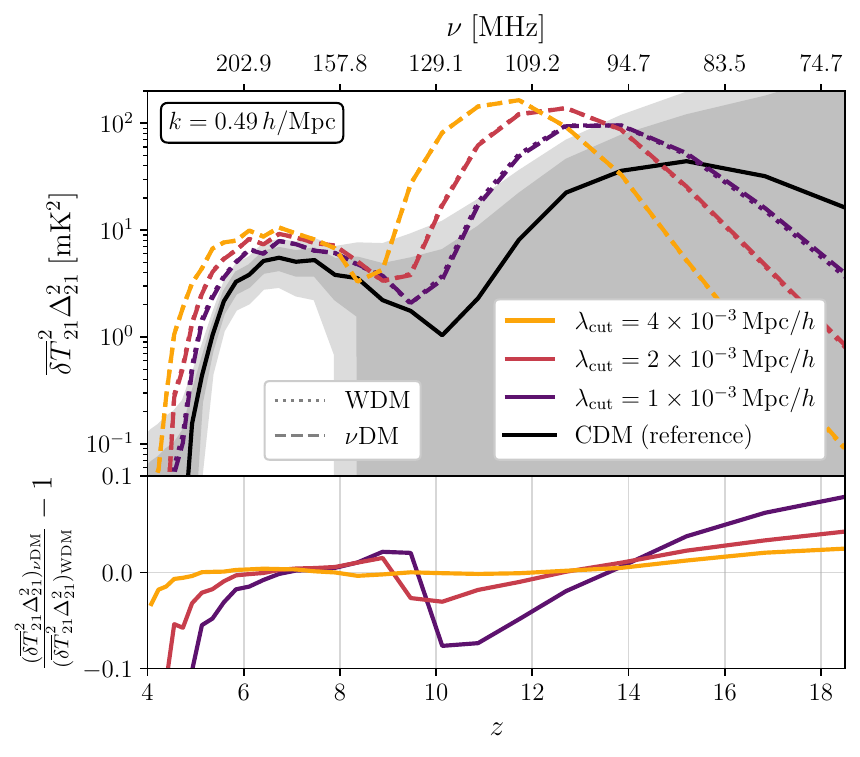}
    \caption{{\bf Upper panels.} $21\,$cm power spectra  as a function of the redshift computed with WDM (dotted) or $\nudm$ (dashed) as the input method for three different values of $\lcut$ at $k=0.20\;h/\rm Mpc$ (left) and $k=0.49\;h/\rm Mpc$ (right). The upper panel of the left plot can be compared to the upper panel of Fig.~\ref{fig:astro-impact-PS-1} assuming the fitting function as an input. Note that the dashed and the dotted curves lie almost exactly on top of each other. {\bf Lower panels.} Ratio of the $\nudm$ and WDM power spectra for different values of $\lcut$.}
    \label{fig:21cm_pk_nudm_WDM_ratio_plot}
\end{figure}

The impact of using WDM or $\nudm$ as the input (method~\ref{it:WDM} or~\ref{it:nuDM}) on the $21\,$cm power spectrum is shown in Fig.~\ref{fig:21cm_pk_nudm_WDM_ratio_plot} for the same values of $\lcut$  as in the Fig.~\ref{fig:astro-impact-PS-1}. Here in particular we show the power spectra  for two different values of $k=0.20\;h/\rm Mpc$ (left, as in the top panels of Figs.~\ref{fig:astro-impact-PS-1} and \ref{fig:astro-impact-PS-2}) and $k=0.49\;h/\rm Mpc$ (right plot). The differences are slightly  stronger for larger values of $k$.

\begin{figure}[t]
    \centering
    \includegraphics[width=0.65\linewidth]{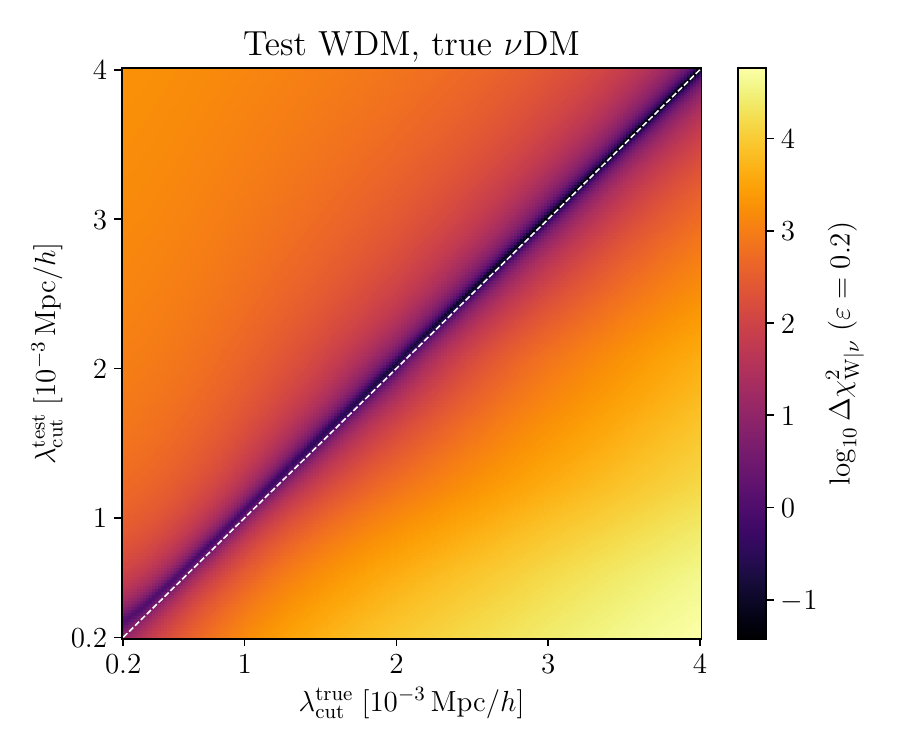}
    \caption{Values of $\Delta\chi^2_{\rm W|\nu}$, testing the goodness of fit of WDM against data given by $\nudm$ as the "true" model shown as a colour gradient as a function of $\lcut^{\rm true}$ and $\lcut^{\rm test}$. For this plot, the modelling error is set to $\epsilon=0.2$.  The dashed white diagonal line indicates where $\lcut^{\rm true}=\lcut^{\rm test}$.}
    \label{fig:AIC_2D}
\end{figure}

The value of $\Delta\chi^2_{\rm W|\nu}$, testing the goodness of fit of WDM as "test" model against data given by $\nudm$ considered to be the "true" model, are  plotted with a gradient of colours in the two dimensional plane of  ($\lcut^{\rm true}$, $\lcut^{\rm test}$) in Fig.~\ref{fig:AIC_2D}. Darker colours indicate the lower $\Delta\chi^2_{\rm W|\nu}$, while  the dashed white diagonal line highlights where $\lcut^{\rm true}=\lcut^{\rm test}$. We show the case where $\varepsilon=0.2$, however setting $\varepsilon=0$ effectively only changes the scaling of the legend, such that $\log_{10}\Delta\chi^2(\varepsilon=0)\simeq1.24\times\log_{10}\Delta\chi^2(\varepsilon=0.2)$. Note that $\Delta\chi^2_{\rm W|\nu}$ is shown as a function of  $\lcut^{\rm test}$ for three values of $\lcut^{\rm true}$ in Fig.~\ref{fig:AIC1D}. As in the latter case, we find that the minimum value of $\Delta\chi^2$ for a fixed $\lcut^{\rm true}$ corresponds to a value of $\lcut^{\rm test}$ slightly below $\lcut^{\rm true}$. This is particularly visible $\lcut^{\rm true}\lesssim1\times10^{-3}\,{\rm Mpc}/h$. This implies that the assumption of the wrong model leads to a small bias in the reconstruction of $\lcut$.

\section{Fisher matrix analysis  and results}
\label{app:fisher}
\subsection{Change of variables with Fisher matrices}
\label{app:fisher_change_variable}

Assuming that the Fisher matrix saturates the Cramer-Rao bound, we may estimate the sensitivity of an observable to a given parameter by determining which value of that parameter is equal to the associated inverse Fisher matrix element (for a 1$\sigma$ detection threshold). In this appendix we detail while this approach must be performed with extreme caution. \\

Consider a model defined by a set of parameters $\btheta$ that can be mapped to another set of parameters $\brho$ by the bijective function $g$ such that $g(\btheta) = \brho$ and $g^{-1}(\brho) = \btheta$. To this model we can associate two Fisher matrices to the same observable. Let us call them $F_{\btheta}$ and $F_{\brho}$ respectively associated with $\btheta$ and $\brho$. From the general expression of a Fisher matrix, our two matrices are nonetheless related by the equation
\begin{equation}
    F_{\btheta} = J^T F_{\brho} J\, , \quad {\rm or~equivalently} \quad F^{-1}_{\brho} = J F_{\btheta}^{-1} J^T\,  ,
\end{equation}
with $J$ the Jacobian of the function $g$ evaluated at the fiducial value. Consider now more specifically the idealised scenario in which our astro/cosmo model is fully described by the analytical transfer function derived in section~\ref{sec:ncdm}. In that case, we can associate a DM mass $m$ to an interaction strength $u$. Let us denote $h(m)$ the value of $u$ that gives the exact same transfer function in the $\nudm$ scenario than a WDM mass $m$ in the WDM scenario. We can then use the previous equations with $g = ({\bf 1}, h)$, $\btheta = (\dots, m)$, and $\brho = (\dots, u)$,  where the dots would denote other model parameters than $u$ and $m$ that would be taken equal in $\btheta$ and $\brho$. It yields
\begin{equation}
    (F^{-1}_{\btheta})_{mm} =  \frac{(F_{\brho}^{-1})_{uu}}{[h'(m^\star)]^2} \, ,
\end{equation}
where we  denote the fiducial values $u^\star$  and $m^\star$ such that $u^\star = h(m^\star)$.
Consider now that the fiducial model is such that 
\begin{equation}
    u^\star = \sqrt{(F_{\brho}^{-1})_{uu}}
\end{equation}
which, for the purpose of this appendix we will call at the \emph{detection} limit of the $\nudm$ scenario. The definition of detection threshold slightly differs in the main text as we account for the fact that the posterior of $u$ has to be truncated and is only defined on positive values, but the exact same conclusion would hold. This simple requirement implies that
\begin{equation}
    \sqrt{(F_{\btheta}^{-1})_{mm}} = \frac{h(m^\star)}{|h'(m^\star)|} 
\end{equation}
which is equal to $m^\star$ if and only if $h(m) = cm^{q}$ with $q = \pm 1$ and $c$ a constant. In other word, unless the the relationship between $u$ and $m$ is linear or inverse there is no way that $\sqrt{(F_{\btheta}^{-1})_{mm}} = m^\star$ when $u^\star = \sqrt{(F_{\brho}^{-1})_{uu}}$. In the specific case where $f(m) = am^b$ (which is the case here, with $b\simeq-2.5$) one has
\begin{equation}
    \sqrt{(F_{\btheta}^{-1})_{mm}} = \frac{m^\star}{|b|} \, .
\end{equation}
That is, for $|b| > 1$ it yields $ \sqrt{(F_{\btheta}^{-1})_{mm}} < m^\star$ and when we detect $u^\star$ we have already detected the corresponding $m^\star$. Said differently, the sensitivity on $m$ that we estimate with the Fisher method is inherently stronger than that on $u$ whatever we do. Notice that this does not depend on the sign of $b$ therefore, considering $1/m$ as a parameter instead of $m$ would not change anything. This result is associated to the fact that, even if one of the two posterior distributions for $u$ or $m$ is Gaussian, then the other can only be Gaussian too if and only if $f(m) = am$.

\subsection{Computing the Fisher matrix: finding the plateaux}
\label{app:fisher_plateau}

Evaluating the Fisher matrix elements requires the derivatives of the $21\,$cm power spectrum over the parameters of the model. Numerically, for a parameter $\theta$ it can be obtained by the evaluation of the power spectrum at two values $\theta_{\pm} = \theta_{\rm fiducial} \pm \delta_\theta$ around the fiducial. For accuracy the step $\delta_\theta$ must be large enough so that the difference of power spectra from $\theta_+$ and $\theta_-$ is not dominated by numerical noise, and small enough so that the two-points estimation of the derivative makes sense. A good guess is usually to consider $\delta_\theta$ to be a few percent of the standard deviation $\sigma_{\theta}$. For all astrophysical parameters we use the values obtained in Ref.~\cite{Mason:2022obt}. However, for an arbitrary cutoff scale, we do not know in advance the value of the standard deviation and must rely on a shooting method. In practice we test a large panel of values for $\delta_\theta$ and try to identify a plateau where the value of $\sigma_{\theta}$ obtained becomes independent of $\delta_\theta$, as it should be. However, this technique is often delicate to apply, due to some residual numerical noise that makes the identification of a single plateau value arbitrary to some level. In that regard, we have developed a fully automatised method in order to overcome the arbitrariness and estimate the uncertainty in the identification of the plateau. 

Let us assume that we have $\{\sigma_\theta^i\}$ for a range of steps $\{\delta_\theta^i\}$ with $i \in [1, N]$. We introduce the 2D \emph{phase-space} vectors ${\bf y}^i=(\sigma_\theta^i, \delta \sigma_\theta^i)$ where $\delta \sigma_\theta^i$ is the gradient of $\sigma_\theta^i$ evaluated as 
\begin{equation}
    \delta \sigma_\theta^i = 
    \begin{cases}
    \displaystyle \frac{\sigma_\theta^{i+1} - \sigma_\theta^{i-1}}{2} \quad & {\forall} i \in [2, N-1]\\
    \sigma_\theta^{2} - \sigma_\theta^1 \quad & {\rm for} ~ i = 1\\
    \sigma_\theta^{N} - \sigma_\theta^{N-1} \quad & {\rm for} ~ i = N\, .
    \end{cases}
\end{equation}
A plateau corresponds to a cluster of ${\bf y}$ values close to $(\mu, 0)$ with $\mu$ a constant. In order to identify such a cluster we try to fit it with a Gaussian distribution of mean ${\bf m} = (\mu, 0)$ and covariance matrix $\mathcal{C}$. Consider that the points in the cluster follow a distribution $\sim P$ with probability density function $p$. Then, we want to minimise the Kullback-Leibler divergence,
\begin{equation}
    D_{\rm KL}(P \, || \, Q) = \int {\rm d}^2 {\bf y} \,  p({\bf y}) \ln \frac{p({\bf y})}{p_N({\bf y} \, | {\bf m}, \mathcal{C})} \, .
\end{equation}
where $p_N$ is the probability density function associated to a bivariate normal distribution. This minimisation problem is equivalent to the minimisation of the cost function
\begin{equation}
    \mathcal{L}_N({\mu}, \mathcal{C}) \equiv - \int {\rm d}^2 {\bf y} \, p({\bf y}) \ln \left[ p_N({\bf y} \, | {\bf m}, \mathcal{C})\right] \sim - \sum_{{\bf y}} \ln \left[ p_N({\bf y} \, |\, {\bf m}, \mathcal{C})\right] \, .
\end{equation}
However, as such, the Gaussian distribution tries to accommodate for points that are far from the plateau and therefore not in the cluster. A solution is, instead of a normal distribution, to use an unnormalised truncated Gaussian distribution defined as
\begin{equation}
    q({\bf y} \, | {\bf m}, \mathcal{C}, r, \lambda ) = 
    \begin{cases}
        p_N({\bf y} \, | \, {\bf m}, \mathcal{C}) \quad & {\rm if} ~ \sqrt{({\bf y-m})^{\rm T} \mathcal{C}^{-1} ({\bf y-m})} \le r \\
        \lambda \quad & {\rm if} ~ \sqrt{({\bf y-m})^{\rm T} \mathcal{C}^{-1} ({\bf y-m})} > r 
    \end{cases}
\end{equation}
with $r$ a chosen value and $\lambda$ a constant chosen such that $\lambda \gg N \exp(-r^2)/\sqrt{|{\rm det}\,  \mathcal{C}|}/(2\pi)$. As a matter of fact, the value of $\lambda$ is not important as long as it is satisfies the previous condition. This choice is such that it will look for a cluster with an extension of $r$ sigmas and a distribution of point that is random elsewhere. In other words, the parameter $r$ quantify the compactness of the cluster making a plateau, and is the only arbitrary input of this method. 

In the end, finding a plateau amounts to minimising
\begin{equation}
    \mathcal{L}_q({\mu}, \mathcal{C} \, | \, r, \lambda ) \equiv -\sum_{{\bf y}} \ln \left[ q({\bf y} \, |\, {\bf m}, \mathcal{C}, r, \lambda)\right] \, ,
\end{equation}
for given values of $r$ and $\lambda$.  In this work we choose $r = 2.58$ (such that the Gaussian covers 99\% of the cluster) and verify a posteriori the validity of our choice for $\lambda$. The minimisation is performed using the stochastic gradient descent algorithm {\tt AdamW} implemented in {\tt pyTorch}. Interestingly, this method automatically gives an estimated error on $\sigma_\theta$. We choose to vary $\lambda_{\rm cut}$ by $20$ linearly distributed values of $\delta_{\lambda_{\rm cut}}$ between 1\% and 5\%.

\subsection{Triangle plots}
\label{app:full_triangle_plots}

\begin{figure}
    \centering
    \makebox[\textwidth][c]{
        \includegraphics[width=1.3\textwidth]{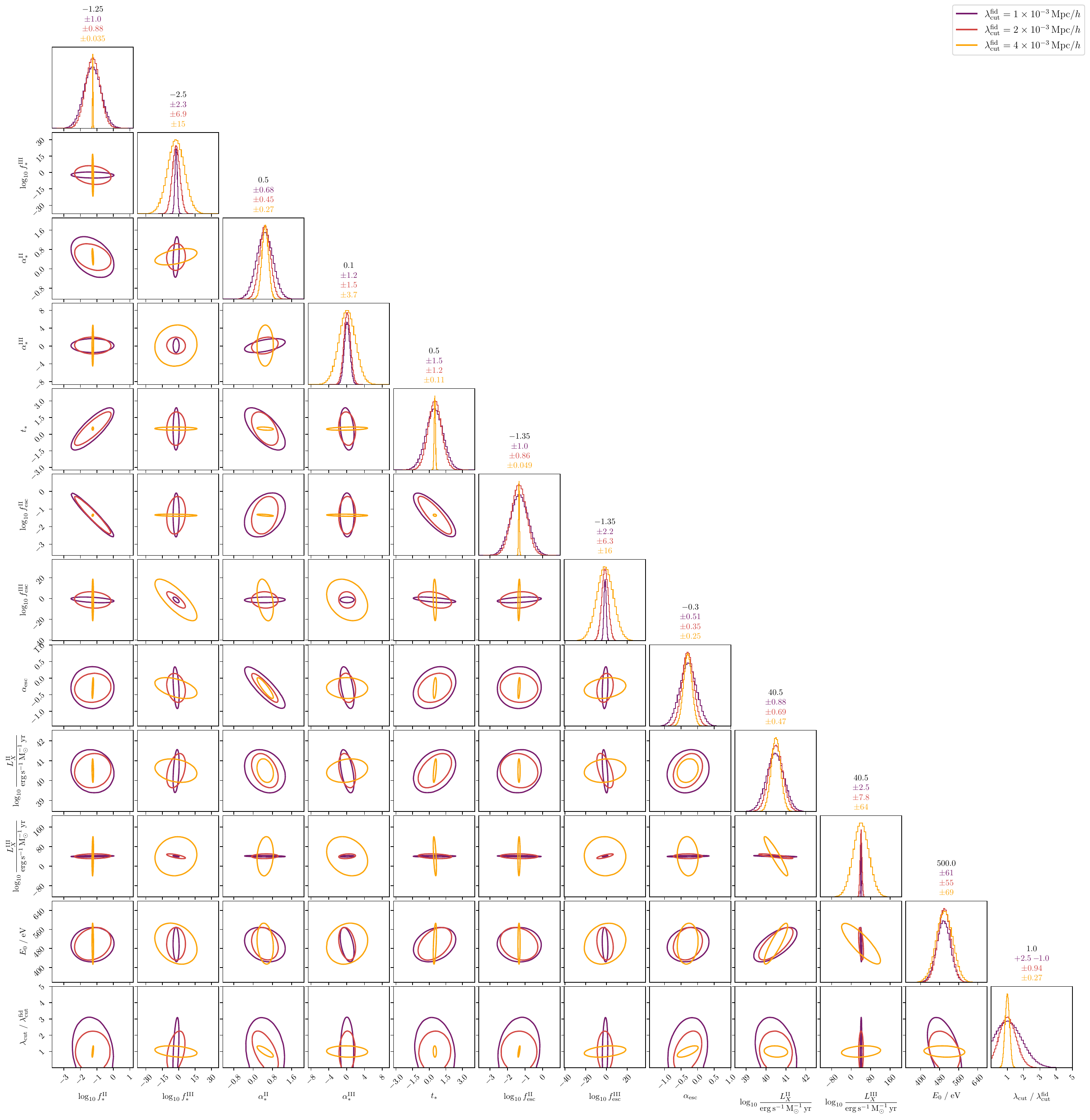}
    }
    \caption{Same as Fig.~\ref{fig:wdm_triangle_plot_small}, but showing all astrophysical parameters and $\lcut$.}
    \label{fig:wdm_full_triangle_plot}
\end{figure}

Fig.~\ref{fig:wdm_full_triangle_plot} shows the marginalised one- and two-dimensional posteriors at $95\%$ CL of the full set of astrophysical parameters considered in the analysis. A selection of degeneracies relevant for the $\lcut$ analysis are already discussed in Sec.~\ref{sec:degen-results}.  Also note that in the colder DM case (purple), we recover similar degeneracies as the one obtained in the Fisher matrix forecasts of Ref.~\cite{Mason:2022obt}, such as the anti-correlation betweens $f_{\rm esc}$ and $f_\star$ and between $\alpha_\star$ and $\alpha_{\rm esc}$, as well as similar 1D uncertainties. Some differences arise due to the fact that the astrophysics and noise models are not exactly the same (\cite{Mason:2022obt} considered an older system temperature description). For example, \cite{Mason:2022obt} uses one single X-ray amplitude which shows similar degeneracies as our MCG amplitude $L_X^{\rm II}$ but not necessarily with $L_X^{\rm III}$.

\acknowledgments
The authors would like to thank J.~Foster, S.~Palomares and B. Zaldivar for insightful discussions. LLH, GF and JS are
 supported by the Fonds de la Recherche Scientifique F.R.S.-FNRS through a senior research
associate, a postdoctoral researcher and research fellow position, respectively. All the authors also acknowledge the support of the FNRS research
grant number J.0134.24, the ARC program of the Federation Wallonie-Bruxelles and the IISN
convention No. 4.4503.15. Computational resources have been provided by the Consortium
des Équipements de Calcul Intensif (CÉCI), funded by the Fonds de la Recherche Scientifique de Belgique (F.R.S.-FNRS) under Grant No. 2.5020.11 and by the Walloon Region
of Belgium. LLH, GF and JS are also members of BLU-ULB (Brussels Laboratory of the Universe, blu.ulb.be).

\bibliographystyle{utphys.bst}
\bibliography{bibliography.bib}

\end{document}